\documentclass[twocolumn,preprintnumbers,amsmath,amssymb]{revtex4-1}
\usepackage{epsfig}
\usepackage{color}
\usepackage{amsmath}
\usepackage{multirow}
\usepackage{float}

\begin{document}

\title{Minimum in the pressure dependence of the interfacial free energy between ice Ih and water}
\author{P. Montero de Hijes$^{1}$, J. R. Espinosa$^2$, C. Vega$^2$, and C. Dellago$^{1,*}$}

\affiliation{$^1$Faculty of Physics, University of Vienna, A-1090
Vienna, Austria}

\affiliation{$^2$Departamento de Qu\'{\i}mica F\'{\i}sica,
Facultad de Ciencias Qu\'{\i}micas, Universidad Complutense de Madrid,
28040 Madrid, Spain}

\begin{abstract}

Despite the importance of ice nucleation, this process has been barely explored
 at negative pressures. 
 Here, we study homogeneous ice nucleation in stretched water   
    by means of Molecular Dynamics Seeding simulations using 
    the TIP4P/Ice model.
    We observe that the critical nucleus size,  interfacial free energy,
            free energy barrier, and  nucleation
            rate
    barely change  between  isobars from
      -2600 to 500 bar  when they are represented as a function of supercooling.
       This allows us to identify universal empirical expressions 
         for homogeneous ice nucleation in the pressure range 
           from -2600 to 500 bar.      
            We show that this universal behavior arises  
            from the pressure dependence
             of the interfacial free energy which we compute by means of the 
              mold integration technique finding
              a shallow minimum around -2000 bar.
             Likewise,  we show that the change
                in the interfacial free energy with pressure
                is  proportional to the excess
                 entropy 
                 and  the slope of the melting line, exhibiting the latter
                  a reentrant behavior also at the same negative pressure.
    Finally, we estimate the excess energy
      and the excess entropy of the ice Ih-water
       interface.

\end{abstract}

\maketitle
$^*$christoph.dellago@univie.ac.at

\section{Introduction}

 Water crystallization is an essential phase transition in nature and technology. 
  However, in the cryopreservation of biological  samples  
 \cite{geidobler2013controlled,xue2015quantifying},
  ice formation can be disastrous.
 The low temperature preserves the biological material
 but causes the water within the sample to be in a metastable state subject to crystallization 
 \cite{pegg2009principles}. Interestingly, by keeping the sample under
  high pressure, ice nuclei are less likely to form keeping 
  water 
   liquid for a longer time \cite{kanno1975supercooling,kalichevsky1995potential,martino1998size}.
   The frequency of the nucleation process is mainly determined by the thermodynamic driving force 
    and the cost of creating the interface between the emerging nucleus and the metastable liquid. 
   The reason  why high pressure slows down the nucleation process is 
    that the difference in chemical potential between ice and water, $\Delta \mu$, 
    which represents the thermodynamic driving force, 
    barely changes between isobars whereas the cost of creating the interface
     notably increases \cite{espinosaPRL2016}. The interfacial free energy is the variable that 
      quantifies this cost. At coexistence, through a planar interface, 
       the interfacial free energy $\gamma_{m}$ differs from the value for
        a critical nucleus $\gamma$ due to the curvature of the surface \cite{tolman1949effect}.
   Nevertheless, both notably increase at high pressure \cite{espinosaPRL2016}.\\

 Homogeneous ice nucleation at standard and high pressure has 
  been extensively explored  
\cite{sanz2013homogeneous,koop2000water,espinosa2014homogeneous,niu2019temperature,espinosa2016time,espinosaPRL2016,piaggi2022,li2011homogeneous,laksmono2015anomalous,amaya2018ice,jeffery1997homogeneous,cheng2018theoretical}.
However,  ice nucleation in 
  water under negative pressure, i.e. stretched water, has
  caught less attention  
  \cite{marcolli2017ice,bianco2021anomalous,rosky2022homogeneous}.
  This process is relevant in porous media
     containing water solutions \cite{roedder1967metastable} and also in  water transpiration inside  plants \cite{wheeler2008transpiration} where negative pressure  occurs.
   Creating and maintaining negative pressure over a sample 
    is non-trivial in experiments \cite{caupin2012exploring}. 
       This is because a liquid  at negative pressure is metastable  
 \cite{debenedetti2021metastable,imre2007existence}.
       In general, this metastability is considered with respect to the vapor phase 
 \cite{menzl2016molecular,caupin2006cavitation},
       although at certain conditions,  it can be also metastable  with respect to ice. 
    Some ingenious approaches to create negative pressure in metastable water 
    include the use of a Berthelot tube \cite{henderson1980berthelot}, 
    centrifugation \cite{briggs1950limiting},
    and more recently the use of acoustic waves \cite{caupin2006cavitation,davitt2010equation,caupin2012exploring}.
   In contrast, in computer
     simulations is  straightforward to work under negative pressures.\\

    In this work, we  investigate 
     how ice nucleation properties are affected
     by negative pressure at different degrees of supercooling.
 In fact, we find
      little effect when pressure changes from strongly negative
       to moderately positive. 
       We investigate
      the role of the interfacial free energy since  it is a key property in determining
       the phase behavior of water at high pressure \cite{espinosaPRL2016}. 
       We find that the slope of the melting line is
        crucial  to describe the change with pressure of the interfacial free energy
         which 
          displays a shallow minimum at negative pressure.
    Our study is based on molecular dynamics simulations
     with the TIP4P/Ice model \cite{abascal2005potential} 
     which has been  extensively used to describe ice nucleation \cite{sanz2013homogeneous,espinosaPRL2016,bianco2021anomalous,niu2019temperature}  and growth \cite{weiss2011kinetic,montero2019ice}  as well as in supercooled
water \cite{debenedetti2020second,lupi2021dynamical}.
     In particular, we employ 
     the seeding technique \cite{sanz2013homogeneous,espinosa2016seeding} 
     to study nucleation and the mold integration 
      technique \cite{espinosa2016ice} to measure the interfacial
       free energy at coexistence.  \\

      \section{Simulation methods}

  All simulations have been done with the GROMACS package (4.6.7-version in double precision)
   with the  TIP4P/Ice water model. 
   The simulations are performed in the
   isothermal-isobaric (NpT) ensemble with
   a time step of 2 fs using the Noose-Hoover
thermostat \cite{nose1984unified,hoover1985canonical}  and the
Parrinello-Rahman barostat \cite{parrinello1980crystal}  both with a relaxation time of
0.5 ps. Electrostatic interactions are accounted for via 
 the particle-mesh-Ewald summation
algorithm \cite{essmann1995smooth} with order 4 and a Fourier spacing of 0.1 nm.
The cutoff for the Lennard-Jones and the 
Coulombic interactions is set to 0.9 nm  and long-range corrections to the Lennard-Jones
 part of the potential are included in energy and pressure.\\

To study nucleation we use the seeding technique \cite{bai2005test, knott2012homogeneous,sanz2013homogeneous}
 which involves the combination of molecular dynamics
    simulations and Classical Nucleation Theory (CNT) \cite{kelton2010nucleation,kashchiev2000nucleation}.
    This technique is based on the behavior of
    a critical nucleus which has equal
    probability of growing and melting 
   when surrounded by the metastable phase
    at the critical pressure and temperature. 
   In practice, one inserts a spherical ice-Ih seed
    in metastable water and then keeps track of the
     time evolution of the size of the cluster. 
     One can vary $T$, $p$
      and the seed size in order to find at which conditions a certain  
       nucleus size is critical ($N_{c}$).
 Once 
      $N_{c}$ is known,  CNT is used to find the interfacial free energy 
       $\gamma $, the barrier height $\Delta G_{c} $, and the nucleation rate $J $.
      Our system sizes ranged between ~80000 and 
       ~250000 water molecules in total. 
       The duration of the trajectories is 
        between 40 and 115 ns. \\

        It is important to note 
         that in the Gibbsian description of 
         interfaces, one has two bulk phases separated by a dividing surface. However, 
         there is some arbitrariness in the location of the dividing surface which 
          also affects  to 
           the interfacial free energy $\gamma$ when the interface has curvature \cite{kondo1956thermodynamical,rowlinson2013molecular,troster2012numerical,montero2020young}. 
          Within the CNT framework, the relevant dividing surface
         is the surface of tension \cite{kashchiev2020nucleation,montero2022thermodynamics}. 
          In order to find the surface of tension we employ an empirical approach 
          that has been successfully applied
        in  crystal nucleation for a large variety of  systems \cite{espinosa2016seeding,montero2019interfacial,espinosa2016ice,espinosaPRL2016,bianco2021anomalous,espinosa2017role}. 
      In this approach,  
        the averaged Steinhardt bond order parameter \cite{lechner2008accurate}, 
        $\bar{q_{6}}(T,p)$ is used in combination with 
     the mislabelling
        criterion \cite{sanz2013homogeneous} to identify ice-like and
       water-like molecules. Within a cutoff distance of 3.5 $\text{\AA}$, we obtain
       $\bar{q_{6}}(T,p)$ for each molecule. The molecules with $\bar{q_{6}}(T,p)$ above 
        a certain threshold $\bar{q_{6,t}}(T,p)$ are labeled as ice whereas those
         below are labeled as liquid. This threshold depends weakly on the considered 
         thermodynamic range covering pressures from -2600 to -1000 bar and temperatures
          from 250 to 270 K (see the supplementary material in 
         Ref. \cite{bianco2021anomalous} for the isothermal
          change in $\bar{q_{6,t}}(T,p)$ with pressure). In this work, the value changes between 0.365 
  for the highest temperature and pressure
            to 0.385 for the lowest temperature and pressure. \\
          
          Once $N_{c}$ is known, we employ the CNT equations \cite{kelton2010nucleation,kashchiev2000nucleation} to determine other important
           parameters. The 
          interfacial free energy 
       $\gamma$ is given as
       
       \begin{equation}
           \gamma = \left( \frac{3N_{c}\rho_{ice}^{2} |\Delta \mu|^3 }{32\pi} \right)^{1/3},
       \end{equation}
       
      \noindent where $N_{c}$ is the size of the critical
       nucleus, $\rho_{\rm ice}$ is the number density
        of ice-Ih in the bulk at the metastable conditions at which the nucleus is critical, 
        and $|\Delta \mu |$  is known
       as the driving force to nucleation,
       i.e. the difference in chemical potential
       between the liquid and ice phases at the conditions which cause the nucleus to be critical.
       This property can be
       obtained by  thermodynamic integration along an isobar \cite{vega2008determination},
       
        \begin{equation}
        \bigg |   \frac{\Delta \mu}{k_{\rm B}T}  \bigg | =   \bigg |  \int_{T_{m}}^{T} \frac{1}{k_{\rm B}T^{2}}\left( \frac{H_{ice} }{N_{ice}}  - \frac{H_{w}}{N_{w}}\right)dT  \bigg | ,
       \end{equation}

 \noindent  where $k_{\rm B}$ is the Boltzmann constant, $T_m$ is the
  melting temperature, and $H$ the  enthalpy,  which  can be
   obtained from simulations of bulk ice-Ih 
   and bulk water along the isobar of interest.
   \\
  
 Then, the free energy barrier is given as
 
 \begin{equation}
      \Delta G_{c} = \frac{16\pi \gamma^{3} }{3\rho_{ice}^{2} |\Delta \mu|^2} = \frac{N_c |\Delta \mu|}{2},
 \end{equation}
 
 \noindent  which allows us to obtain the nucleation rate $J$,
 the number of critical nuclei forming per unit of time
   and volume. According to CNT, $J$ is given as
 
 \begin{equation}
     J = \rho_{w} \sqrt{ \frac{|\Delta \mu|}{6\pi k_{\rm B}T N_{c}} } f^{+} \exp \left( -\frac{\Delta G_{c} }{k_{\rm B}T}\right) ,
     \label{eq:lajota}
 \end{equation}
 
 \noindent  where $f^{+}$ is the attachment rate which can be
  approximated through this expression \cite{espinosaPRL2016,bianco2021anomalous}
 
 \begin{equation}
     f^{+}  = \frac{24 D_w N_{c}^{2/3}}{\lambda^{2}} ,
 \end{equation}
 
  \noindent  where $D_w$ is the diffusion coefficient of the
  metastable liquid and $\lambda$ is a characteristic
   length, the typical distance that a water molecule covers in order to attach into the nucleus,
   whose value is approximately 3.8  $\text{\AA}$ for water
   \cite{espinosaPRL2016,bianco2021anomalous}.
    \\

   To find the ice-Ih-water interfacial free energy at coexistence for a planar interface,   $\gamma_m$, 
    we use the
   mold integration  technique \cite{espinosa2016ice},
   which consists in computing the
reversible work $W$ that is necessary to form a crystal slab
within a liquid at coexistence. This work is related to the interfacial free energy at
 coexistence, $\gamma_{m}$, by $W=2A\gamma_{m}$ where $A$ is the interfacial area and
  the number 2 accounts for the two interfaces of the slab.
  The  slab formation
is induced by switching on an attractive interaction between  the 
mold  of potential energy wells and the particles of the initial liquid. 
The wells are arranged in  the equilibrium positions of the oxygen
atoms in the ice facet under investigation at coexistence
conditions, i.e. 
for temperatures and pressures located
along the ice Ih-water equilibrium line for the TIP4P/Ice water model.
 First, one has to obtain $\gamma_{r_w}$,  which is given  as

 \begin{equation}
     \gamma_{r_w} = \frac{1}{2A} \left( \epsilon_{w} N_{w} - \int_{0}^{\epsilon_{w}} \langle N(\epsilon)      \rangle d\epsilon    \right) ,
 \end{equation}
 
  \noindent  where $r_w$ indicates the radius of the potential wells 
  and $\epsilon$ is their 
  energy  (with maximum depth equal to 
  $\epsilon_{w}$).  $N_{w}$ is the number of wells in the mold and $\langle N (\epsilon) \rangle $
   is the average number of occupied wells at a given potential depth $\epsilon$. 
   The integration needs to be reversible. To ensure this, 
  thermodynamic integration is performed for wells whose radius
is larger than a certain value $r_w^0$.  At $r_w^0$ the slab is fully formed and the
  stability no longer depends on the mold-liquid interactions, hence, leading to potentially
   irreversible ice growth. 
   However, since 
   this is the radius that
 recovers the actual value of  $\gamma_m$, 
thermodynamic integration is repeated for several values of
 $r_w < r_w^0$ and then $\gamma_{r_w}$ is extrapolated to its value at
 $r_w^0$ giving $\gamma_m$ \cite{espinosa2016ice}.\\

\section{Results}

\subsection{Universality in ice nucleation variables at negative and moderate pressure}

  First, we study nucleation along the isobars of  -2600, -2000, and -1000 bar by means of the seeding approach.
 For pressures below -3000 bar we observed 
 spontaneous cavitation occurring within the time scale of the trajectories needed in the seeding method.
  We obtain
    the critical nucleus size $N_c$,
     the driving force to nucleation $|\Delta \mu|$,
      the interfacial free energy $\gamma$,
       the free energy barrier to nucleation $\Delta G_{c}$
        and the nucleation rate $J$. These results are
         presented in Table \ref{tab:seeding}. 
As can be seen, even though the pressure  
         significantly differs, the results are surprisingly similar 
         for nuclei of similar size for equivalent supercoolings. 
         This behaviour is considerably different from what has been found when comparing the nucleation scenario of normal vs. high pressure (i.e. 2000 bar; Ref. \cite{espinosaPRL2016}), where the increase in pressure 
          brings down the ice nucleation rate.
         \\

\begin{table*}[]
\begin{ruledtabular}
\begin{tabular}{ccccccccccc}
$N_{c}$    & $T$ [K] & $\Delta T$ [K] & $p$ [bar]& $\rho_{w}$ [g/cm$^3$]& $\rho_{\rm ice}$ [g/cm$^3$] &$|\Delta \mu|$ [kJ/mol] & $\gamma$ [mJ/m$^2$]&  $\Delta G_{c}$ [kJ/mol]  & log$_{10}(J$ [m$^{-3}$s$^{-1}$])\\
\hline
1650  & 255  & 23   &  -1000  & 0.9208 &  0.8999 &  0.367  & 21.62 &  303 & -24 \\
7450 &  264 & 14   &  -1000    & 0.9285  & 0.8985 & 0.237 & 23.03 & 883 & -136\\
\hline
1750   & 255  &  25  &  -2000  & 0.8855 & 0.8916 & 0.367 & 21.87   & 321 & -28 \\
7600    &  266 &  14  &  -2000  &   0.8876 & 0.8894 &  0.224 &  21.74 & 850 & -128\\
\hline
 1950 & 255  & 24  &      -2600  & 0.8674 & 0.8866 & 0.340 & 20.89   & 332 & -30   \\

\end{tabular}
\end{ruledtabular}
\caption{ Seeding results in tabular form. $N_c$ is the critical nucleus size, $T$ and $p$ are
 the thermodynamic conditions that make such nucleus size to be critical, and $\Delta T$ is the
  supercooling, $T_m - T$. The densities of water $\rho_w$ and ice $\rho_{\rm ice}$ are also shown,
   as well as the interfacial free energy at nucleation $\gamma$, the barrier height $\Delta G_c$,
    and the base-10 logarithm of the nucleation rate log$_{10}(J)$.}
\label{tab:seeding}
\end{table*}

               \begin{figure*}[htb!]
\centering
\includegraphics[width=3.1in]{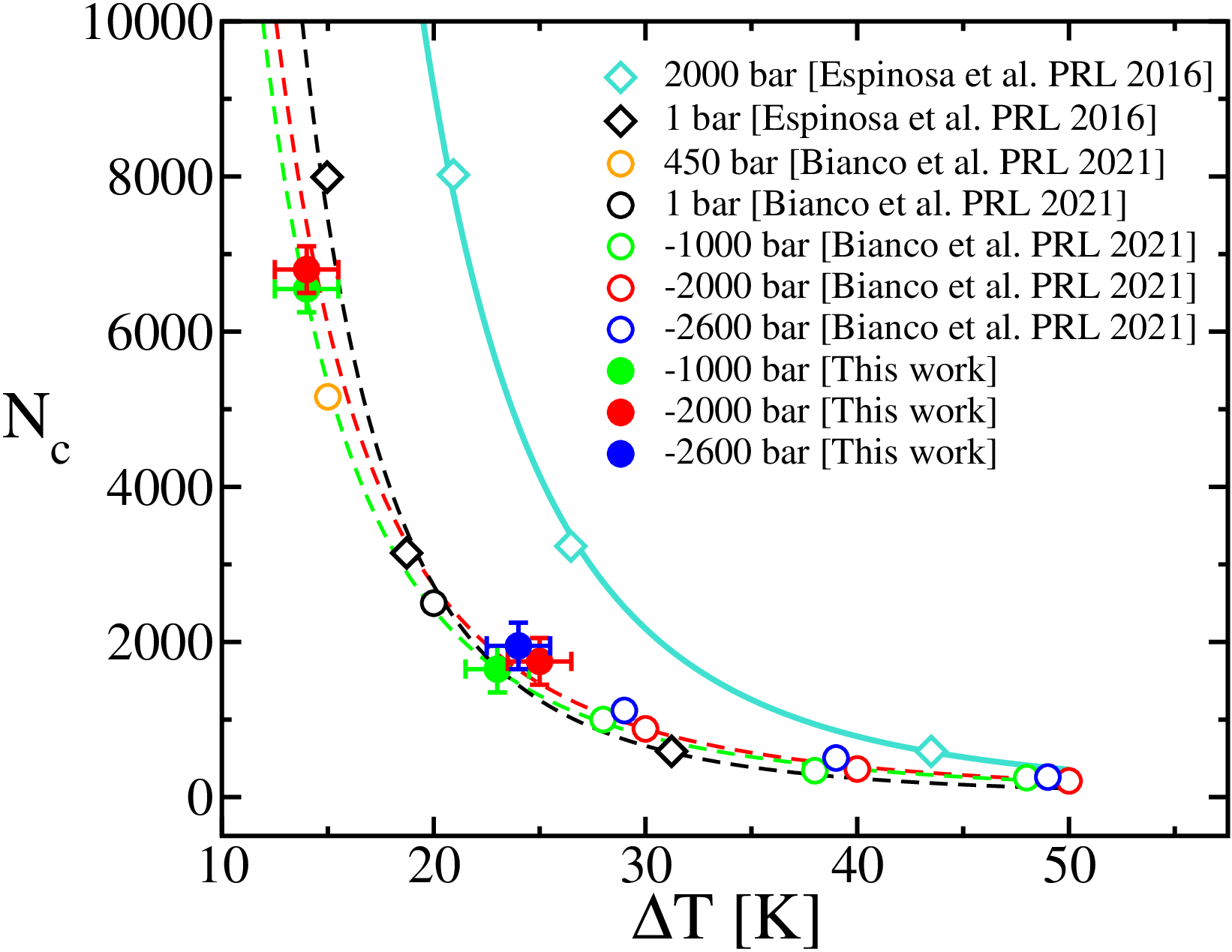}  a)
\includegraphics[width=3.1in]{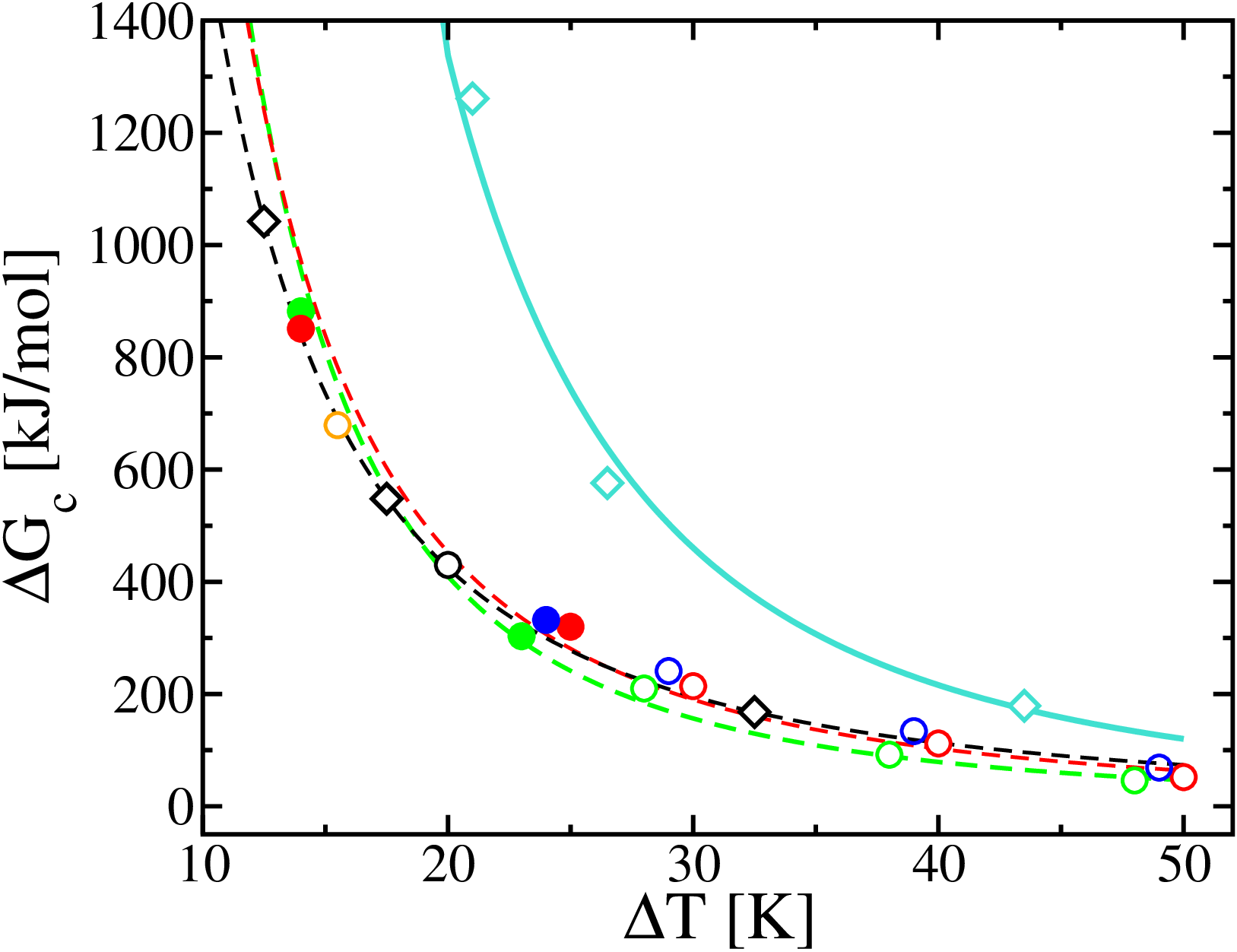}  b)

\caption{\label{fig:nucleation1} a) Critical nucleus size  and b) 
free energy barrier to undergo nucleation against supercooling.
     The same legend applies in both panels. Numerical details can be seen in Table \ref{tab:seeding}.
     The color indicates the pressure, whereas solid symbols correspond
      to  simulations performed in this work, and empty symbols correspond to data obtained
       from previous work as indicated in the legend.
        For the same pressure but different work we use different
        symbols. The  lines are power law fits to points sharing the same pressure
         independently on the work in which they were obtained.}
\end{figure*}

     To further understand this behavior,  we connect our results with
          those from previous works where nucleation
          had been studied for the TIP4P/Ice model 
          at different pressures
           including negative, moderate, and high pressure states \cite{espinosaPRL2016, bianco2021anomalous}.
           In Fig. \ref{fig:nucleation1} a) we show the critical nucleus size
            as a function of supercooling for several isobars. 
            We provide results at moderate supercoolings at -2600, -2000, and -1000 bar.
            For these same isobars as well as for the 1 bar isobar, we show the values reported  in Ref. 
  \cite{bianco2021anomalous}. For the 1 bar isobar, we
             also show the values given in Ref. \cite{espinosaPRL2016}, 
              which
              also provides with the values at the 2000 bar isobar. 
            As can be seen, only the points corresponding to the 2000 bar
            isobar \cite{espinosaPRL2016}
            exhibit a different trend. The isobars at
             -2600, -2000, and -1000 bar from this work as well as from Ref. \cite{bianco2021anomalous}, and the 1 bar isobar from both Refs. \cite{espinosaPRL2016, bianco2021anomalous}
             follow approximately the same curve. 
           Notice  that even a point at 450 bar reported in
             Ref. \cite{bianco2021anomalous} was included being
             in agreement with this group of isobars. 
              In fact, as shown in Fig. \ref{fig:nucleation1} b), pressure
              hardly affects the nucleation free energy barrier
              as a function of supercooling from -2600 bar to 450 bar.\\
               
\begin{figure*}[htb!]
\centering
\includegraphics[width=3.1in]{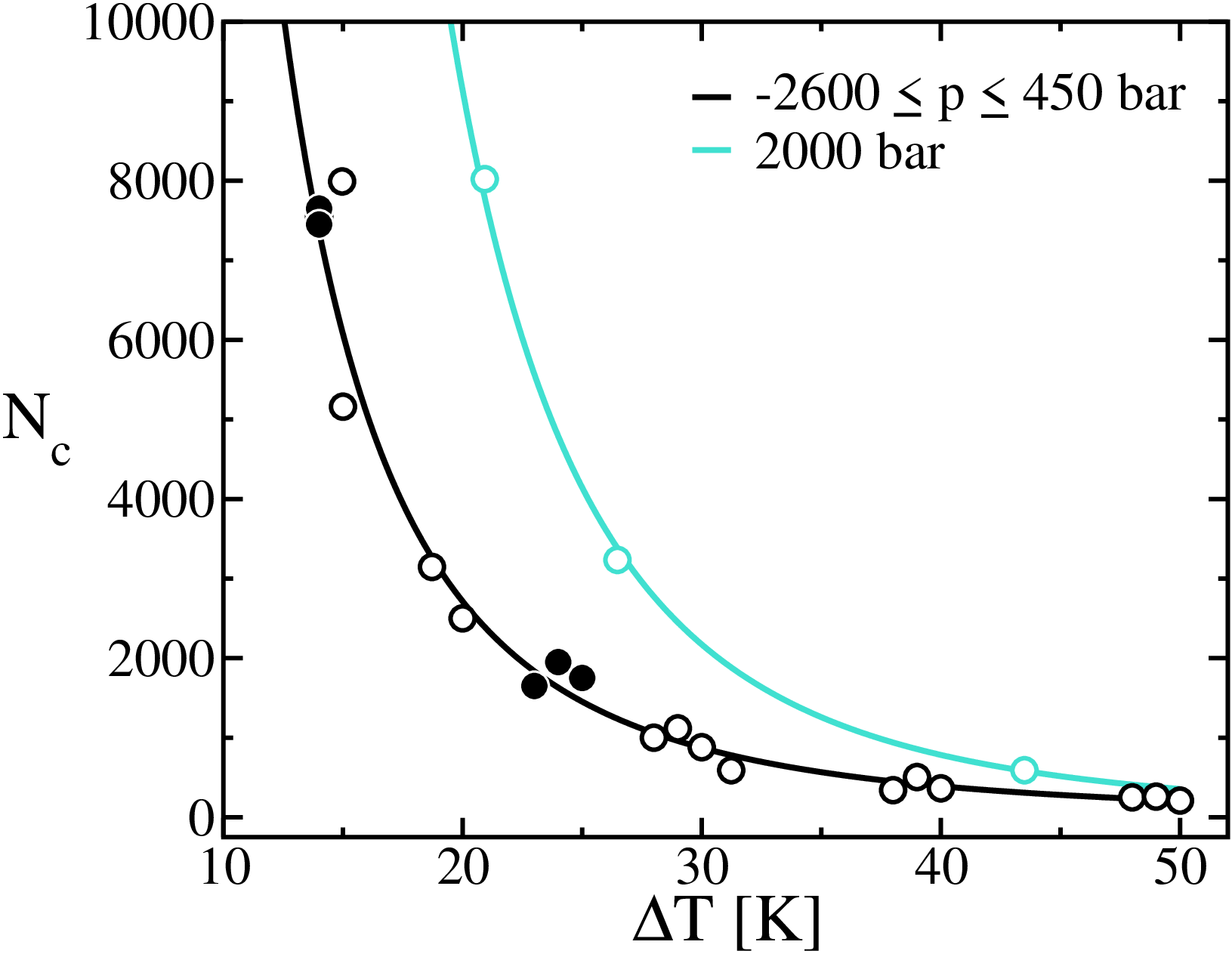}  a)
\includegraphics[width=3.1in]{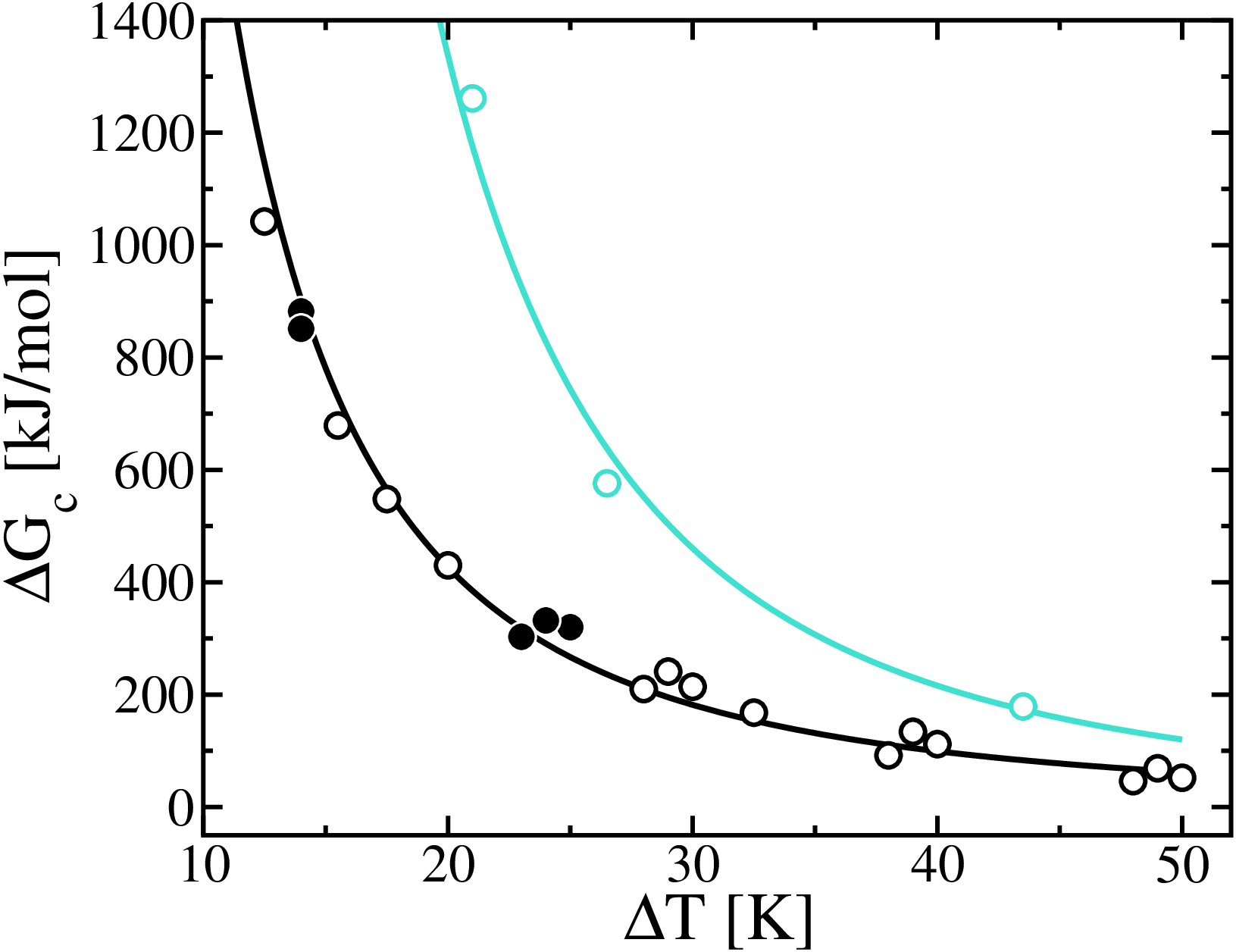}  b)
\includegraphics[width=3.1in]{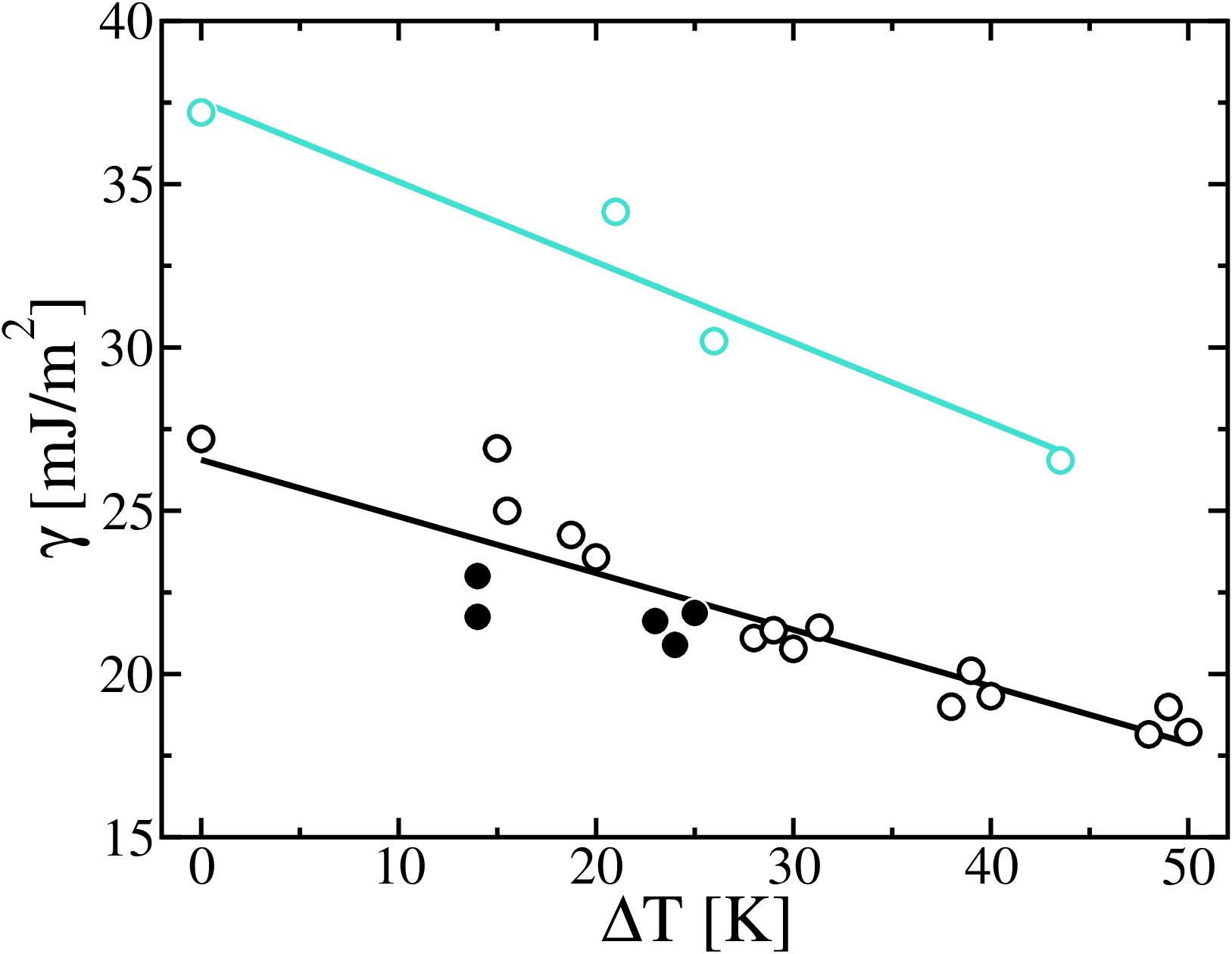}  c)
\includegraphics[width=3.1in]{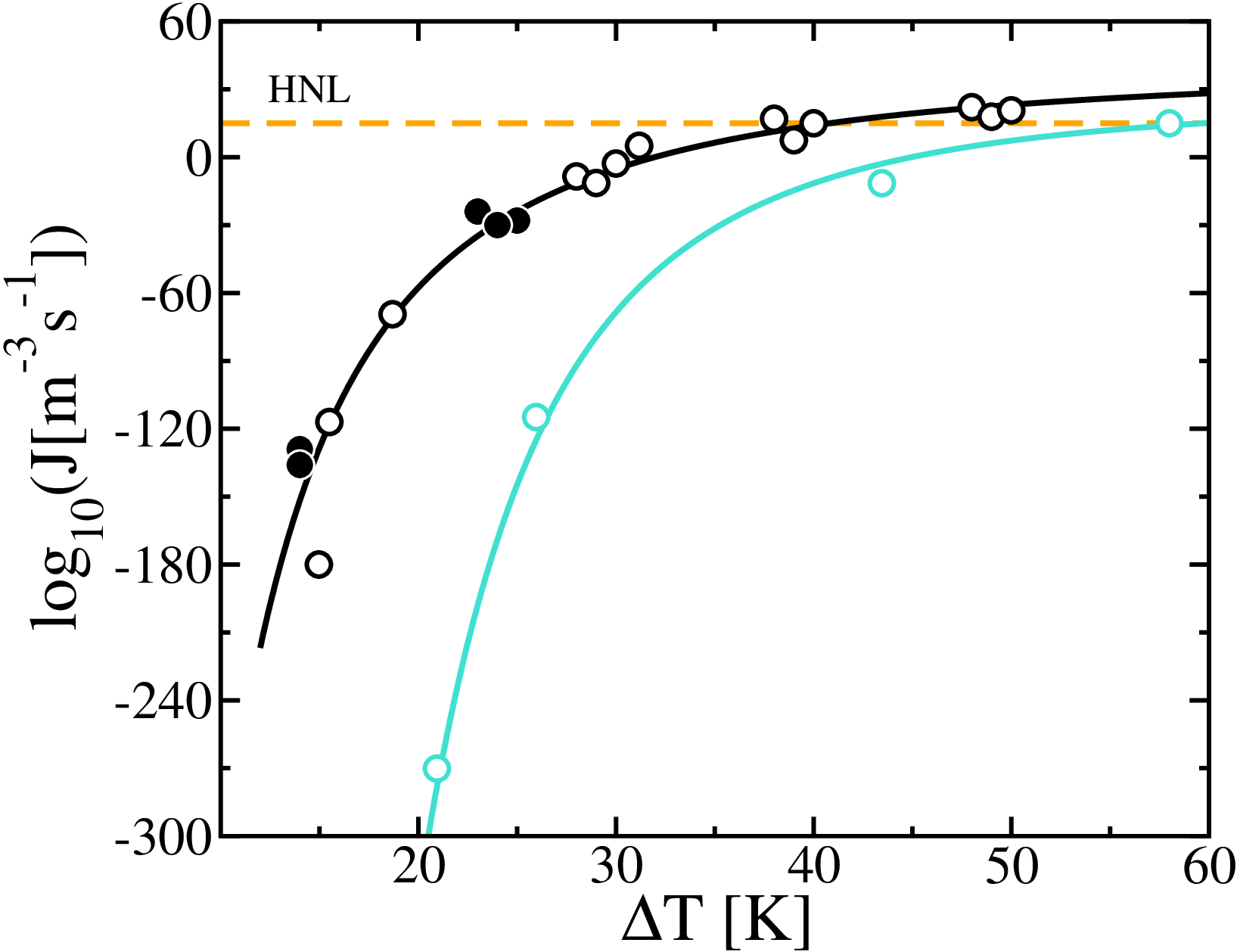}  d)

\caption{\label{fig:nucleation} a) Critical nucleus size,
b)  nucleation free energy barrier, c) interfacial 
 free energy, and d) $\log_{10}J$ against supercooling.
     The same legend applies to all  panels. The color indicates the pressure regime according to 
      the legend. Points obtained
     in this work are shown as solid
     symbols whereas results from Refs. \cite{espinosaPRL2016,bianco2021anomalous} as empty
     symbols. Black solid symbols are restricted to pressures between -2600 and -1000 bar,
      whereas black empty symbols cover from -2600 up to 450 bar. Cyan empty symbols correspond 
       to 2000 bar. For each magnitude, a common fit
      to our data and those of Refs. \cite{espinosaPRL2016,bianco2021anomalous} is included.
      For panels a) and b) a power law fit is used as given by Eq. \ref{eq:Nc} and Eq. \ref{eq:Gc} 
      respectively, whereas 
       for panel c) we use a linear fit (Eq. \ref{eq:commong}) and for panel d)
       we use a CNT-based fit. HNL in panel d) is the iso-nucleation line of log$_{10}$(J[m$^{-3}$s$^{-1}]$) = 15.}
\end{figure*}

               Our results from Fig. \ref{fig:nucleation1} suggest that a similar nucleation behaviour as
               a function of supercooling may take place
               from -2600 to 450 bar. 
               That is a strikingly different behavior to the one observed when 
               increasing pressure  to 2000 bar.
               Thus,  we  propose 
               universal empirical expressions for the variation of 
               different homogeneous ice
               nucleation properties with the 
               supercooling
               independently of the pressure as long as it is within this regime. 
              Nevertheless, we first need to confirm 
              that what was observed for $N_{c}$
               and $\Delta G_{c}$ also applies to $J$.
                 In Fig. \ref{fig:nucleation} 
               we show again a) $N_{c}$ and b) $\Delta G_{c}$ as well as c) $\gamma$ and
                d) $\log_{10}J$. This time, for each
                 magnitude, we include
                 a common fit  to 
               data  from moderately positive to deeply negative pressure (including our own data and those from Refs. \cite{espinosaPRL2016,bianco2021anomalous}) along a separate fit
                at high pressure \cite{espinosaPRL2016}. 
                In 
               c) we show $\gamma$ which exhibits higher variance.
                Finally, in d),  we show how very different pressures
               (from largely negative
                 to moderately positive) 
                 lead to approximately the same nucleation rate $J$ 
                 as a function of supercooling, $\Delta T = T_m - T$. The values of $T_m$
                  are given in Table \ref{tab:gammas}. Hence, 
                 we can use the respective common fit
                  as   universal
                 empirical expressions to describe the change with supercooling within this broad range of pressures.\\

                 For  $N_{c}$ we obtain 
  \begin{equation}
        N_{c}(\Delta T) = 1.2\cdot 10^{7} \cdot \left( \frac{\Delta T}{T_0} \right)^{-2.8}
        \label{eq:Nc}
    \end{equation}
    
  \noindent   and for $\Delta G_{c}$ (in kJ/mol)

    \begin{equation}
        \Delta G_{c} (\Delta T) = 2.3\cdot 10^{5} \cdot \left ( \frac{\Delta T}{T_0} \right)^{-2.1},
        \label{eq:Gc}
    \end{equation}

  \noindent    where $T_0$ equals 1 K for correctness of units.   For $\gamma$ in mJ/m$^2$, we obtain 
  
  \begin{equation}
      \gamma (\Delta T) = 26.6 - 0.174\cdot \Delta T 
      \label{eq:commong},
  \end{equation}

  \noindent  and, finally,  for $J$ in m$^{-3}$ s$^{-1}$, one should use Eq. \ref{eq:lajota} along with Eq. \ref{eq:Gc} (after converting into in $k_{\rm B}T$ units), and 10$^{36}$  m$^{-3}$s$^{-1}$
 as the prefactor
$(\rho_{w} \sqrt{ |\Delta \mu|/(6\pi k_{\rm B}T N_{c})} f^{+} )$.\\

\begin{figure}[h!]
\centering
\includegraphics[width=3.1in]{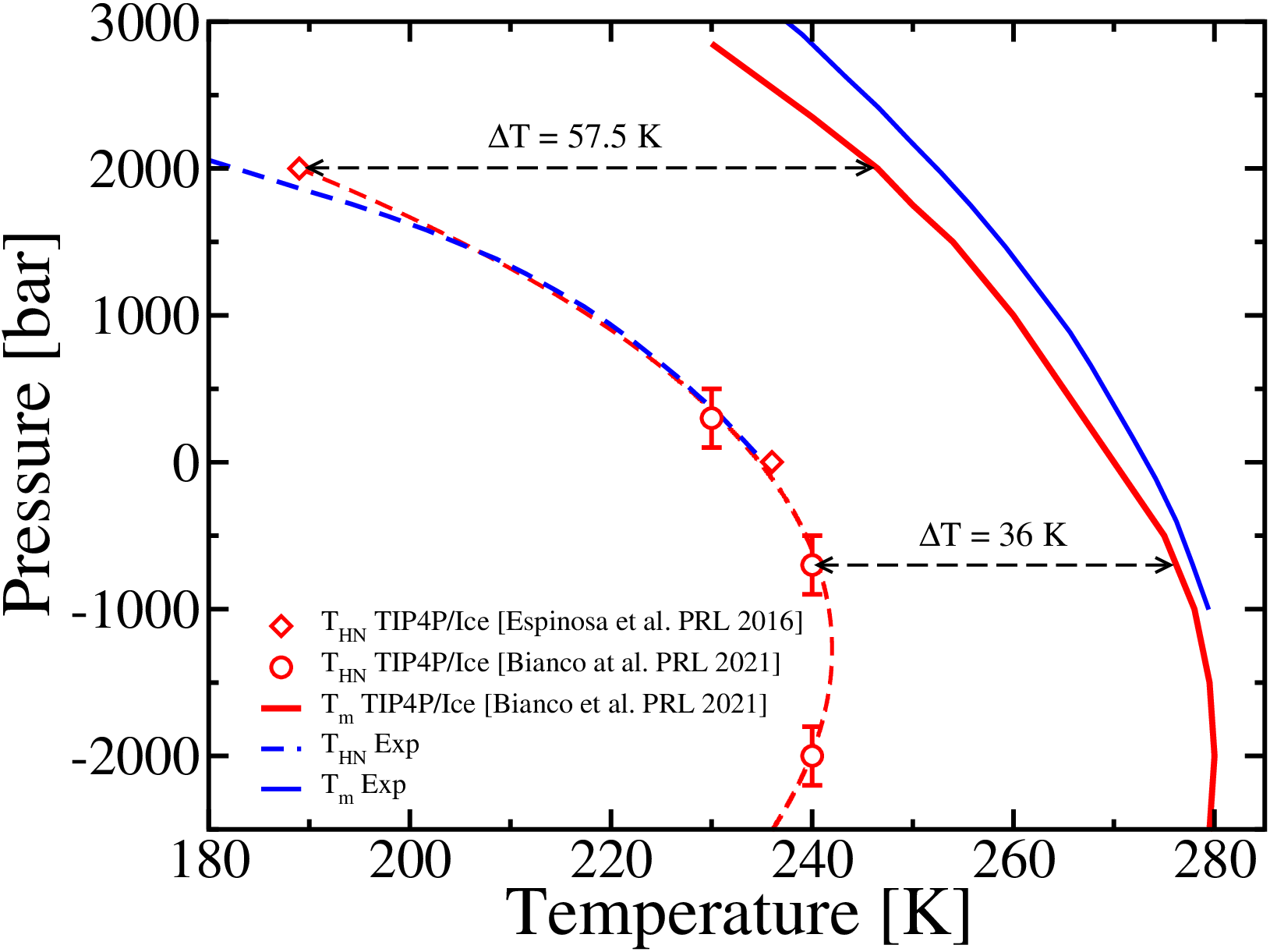}   

\caption{\label{fig:coex} In solid lines, the coexistence lines
 T$_{m}$ where blue is experimental \cite{marcolli2017ice} and red is for
  the TIP4P/Ice \cite{bianco2021anomalous}.
  The dashed blue line corresponds to the experimental HNL \cite{kanno1975supercooling}. Empty red symbols correspond to simulation
  estimates for the TIP4P/Ice of the HNL for $\log _{10} J$/(m$^{-3}$s$^{-1}$) = 15 (the dashed red line is a guide 
  connecting these points). The turning point of the melting curve of 
  TIP4P/Ice occurs at 280K and -2000 bar.}
\end{figure}

 The results shown in Fig. \ref{fig:nucleation} have  interesting consequences. First, taking into account that 
 $N_{c}$ and $\gamma$ (panels a) and c) respectively) are roughly independent of $p$
  when it goes from largely negative to moderately positive pressures, the isobaric Tolman length which determines the change
   in $\gamma$ with the inverse of the radius of curvature of the cluster along an isobar \cite{tolman1949effect,schmelzer2019entropy,baidakov2019spontaneous} 
   is roughly constant too and equal to 0.24(5) nm, where the parenthesis indicates
     uncertainty in the last digit. This result is in agreement with previous work \cite{montero2019interfacial}. 
  Second, in panel
 d) one can see that from strongly negative to 
   moderately positive pressure we obtain the same nucleation rate with respect
    to the supercooling which means that the homogeneous nucleation
     line (HNL) should be at a constant distance to the melting
      line in this regime
     as predicted recently for this water model \cite{bianco2021anomalous} as well
      as for the mW model \cite{molinero2009water} in Ref. \cite{rosky2022homogeneous}. In Fig. \ref{fig:coex}
      we show the estimates for the model \cite{espinosaPRL2016,bianco2021anomalous} assuming that the 
      HNL corresponds to an iso-nucleation rate of $\log _{10} J$/(m$^{-3}$s$^{-1}$) = 15 and we  compare it to the experimental HNL \cite{kanno1975supercooling}. Also, 
      the coexistence lines of the model \cite{bianco2021anomalous} and the experimental one \cite{marcolli2017ice}
       are presented  showing how the distance between
       the coexistence line and the HNL is roughly constant
       until pressure increases enough such that the required supercooling
        to reach $\log_{10}J = 15$ becomes larger.
However, even though this result might be useful,
a physical explanation is still missing.
 In order to answer this question, we look at the
  pressure-induced deceleration of ice nucleation. In 2016,
  Espinosa et al. \cite{espinosaPRL2016} showed that the origin of this
   phenomenon arises from the increase with pressure of 
    the interfacial
   free energy both at coexistence  $\gamma_{m}$ and for nucleation ($\gamma$ at 
    a given supercooling $\Delta T$) while the difference in chemical potential $\Delta \mu$ does not change
     so much with  $\Delta T$.  Thus, one needs a larger $\Delta T$  to obtain
       the same $J$ at high pressure. 
       In this work, we observe approximately the same $J$
       as a function of $\Delta T$ from strongly negative to moderately positive pressure. \\

\begin{figure}[h!]
\centering
\includegraphics[width=3.1in]{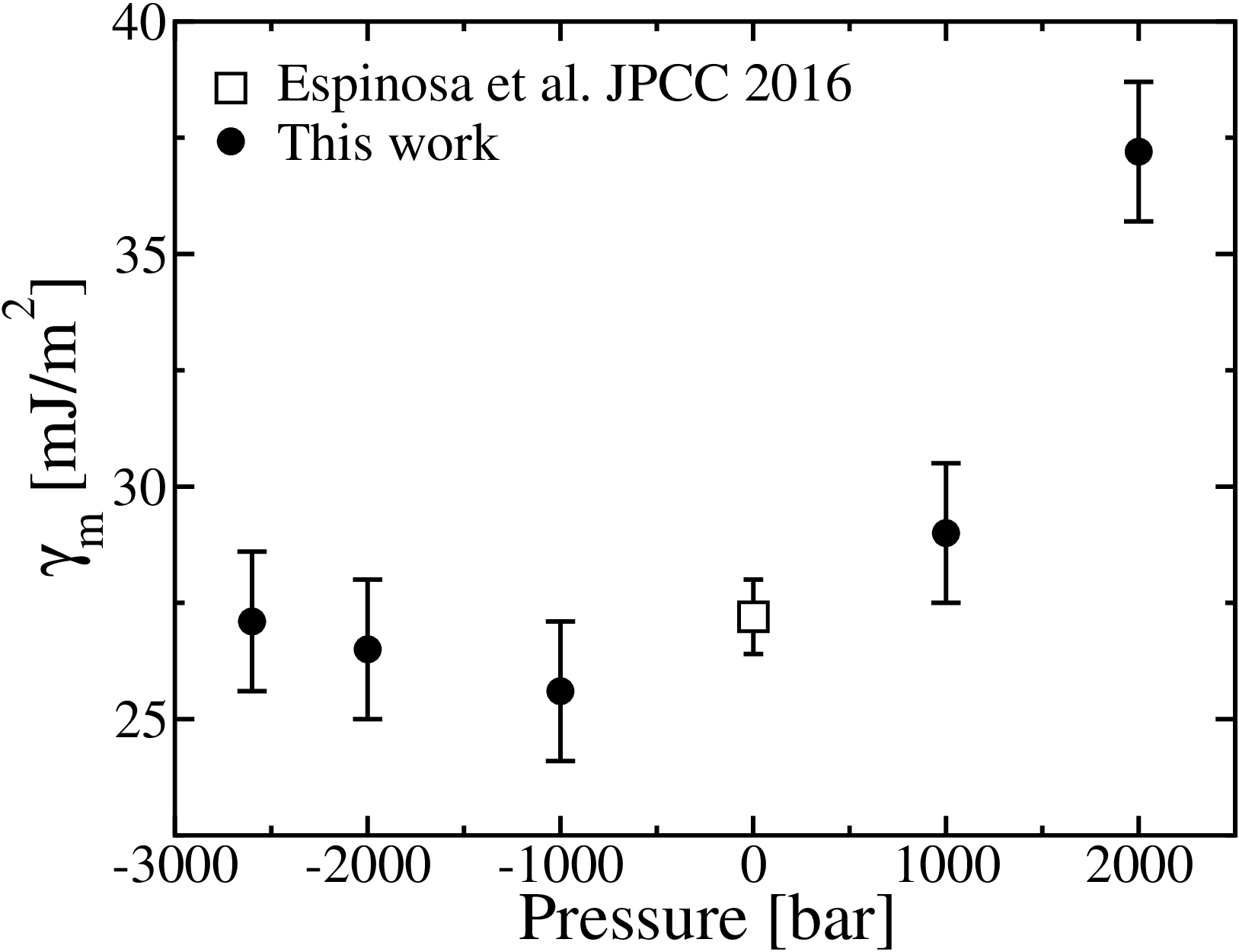} 
\caption{\label{fig:gcoex}  Ice Ih-water interfacial free energy at coexistence for
 the basal plane for the TIP4P/Ice model. 
   }
\end{figure}

\begin{figure}[h!]
\centering
\includegraphics[width=3.1in]{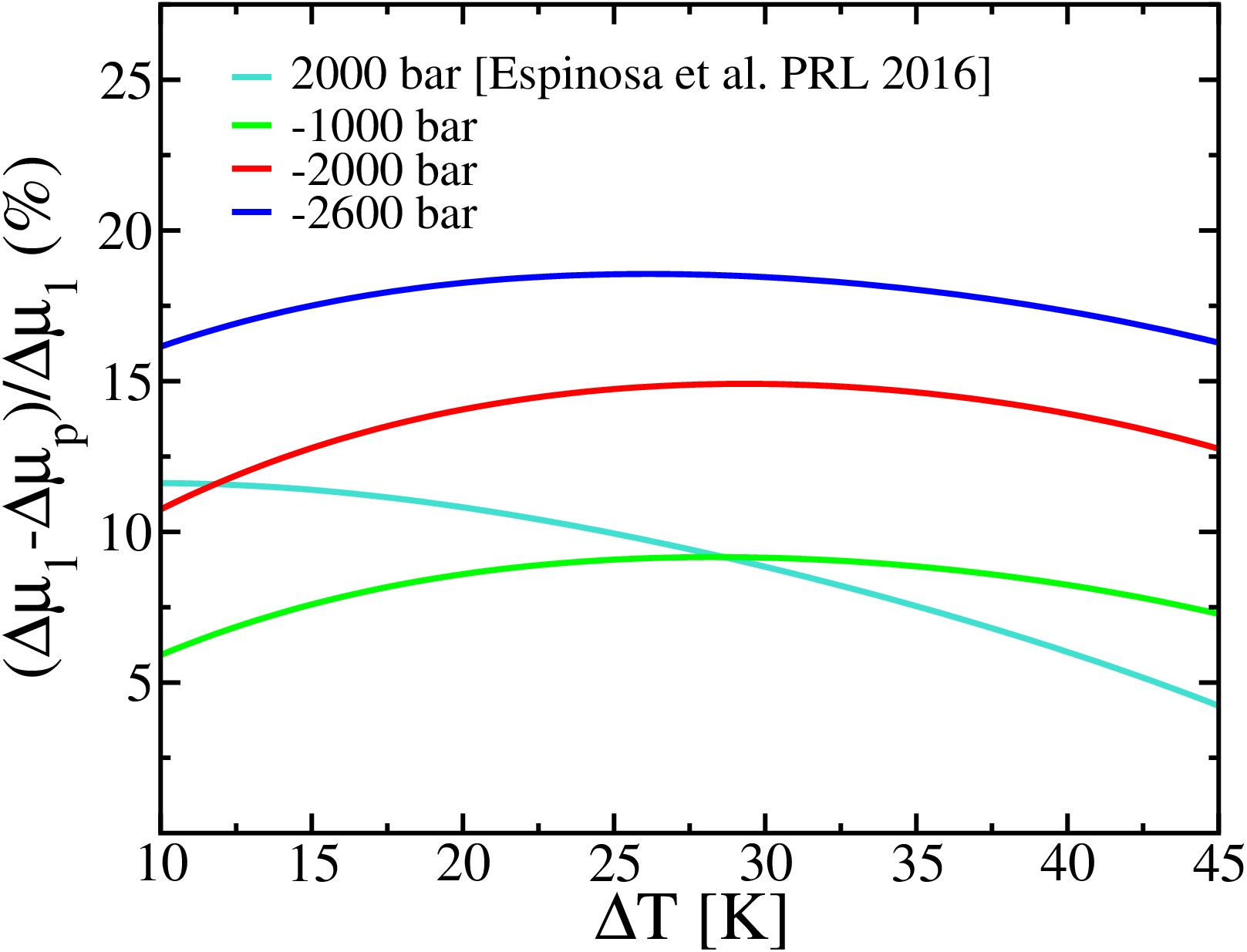} a)
\includegraphics[width=3.1in]{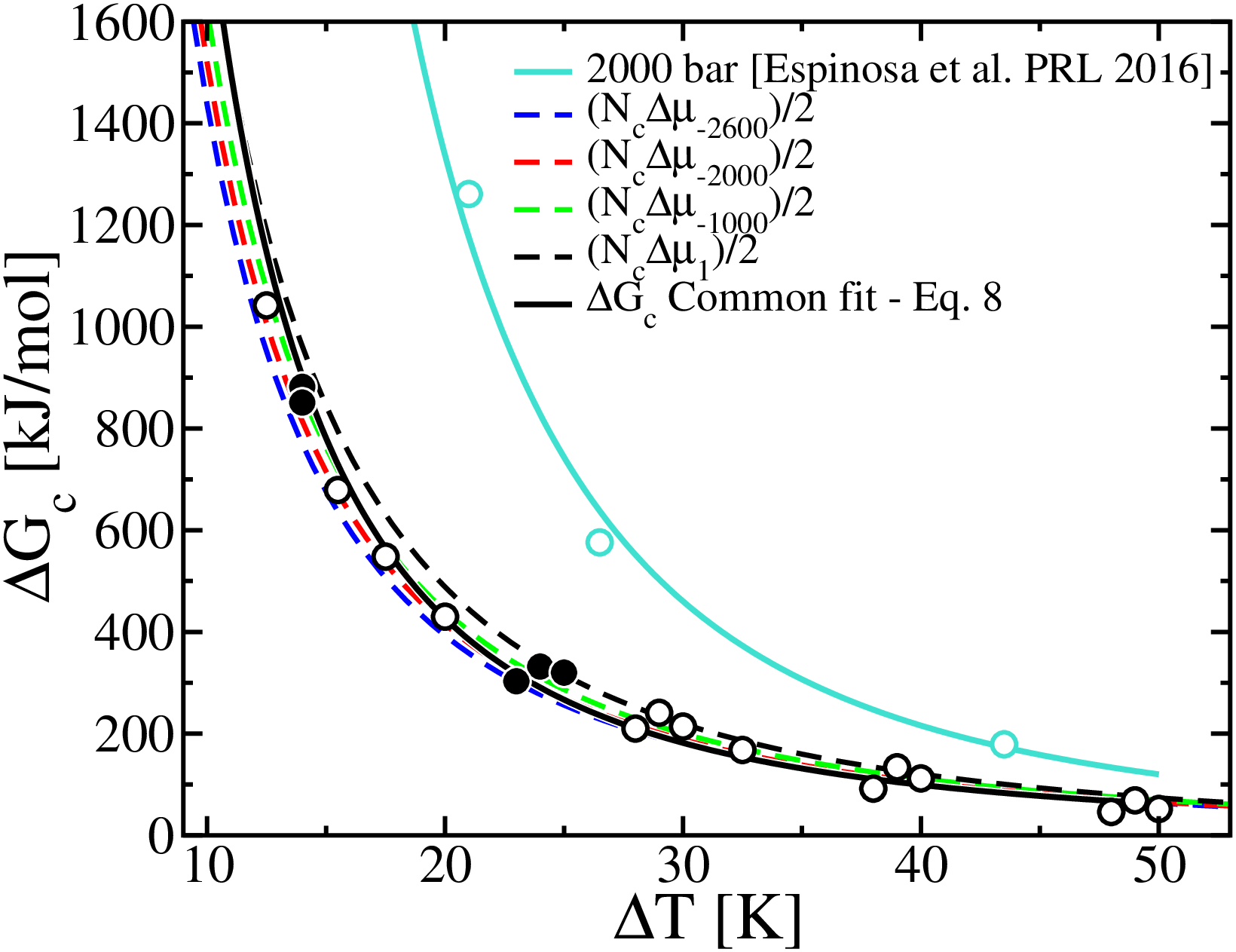} b)

\caption{\label{fig:gcoex2} 
a) Deviation in $\Delta \mu$ at different isobars (-2600, -2000, -1000, and 2000 bar)  with respect
 to the one at 1 bar.
  b)
      In dashed black (for 1 bar), green (for -1000 bar),
      red (for -2000 bar), and blue (for -2600 bar) lines,
       we present 
      free energy barriers $\Delta G_{c} = N_{c}\cdot\Delta \mu/2$ where $N_{c}(\Delta T)$ is given by
       the common fit of Eq. \ref{eq:Nc} and
       for $\Delta \mu$ we use the corresponding
        values for each isobar. In solid black line the 
        common fit for $\Delta G_{c}$ proposed in
         Eq. \ref{eq:Gc} and in turquoise the fit for 2000 bar
          from Ref. \cite{espinosaPRL2016}. As black circles
          we show the data in the -2600 bar $< p <$ 450 bar regime, where solid circles are computed in this work
           and empty come from Refs. \cite{espinosaPRL2016,bianco2021anomalous}.
           }
\end{figure}

  Since we obtain roughly the same $\gamma$ as a function of $\Delta T$ at different
   negative pressures, we expect also $\gamma_{m}$ to barely change
   with $p$. The term $\gamma_m$ refers to a planar interface between 
   ice and water at certain conditions along the coexistence line whereas the term $\gamma$ 
   refers to a curved interface between a critical nucleus of ice  and water at a certain supercooling $\Delta T$ 
    along an isobar. In both cases, thermodynamic equilibrium holds.
   However, when the interface is planar then the pressure is equal in both phases while
   in a spherical interface  the  pressure changes between
      phases following the Young-Laplace equation. Then, we compute  $\gamma_{m}$ for several points.
   In addition to the negative pressure isobars,
    we compute two points at 1000 bar and 2000 bar respectively. 
  We  study only the basal 
  plane as we do not
 expect severe anisotropy (as much as $\sim$ 10$\%$) with the prismatic ones \cite{handel2008direct,davidchack2012ice,rozmanov2012anisotropy,espinosa2016ice}. The results are presented in Table \ref{tab:gammas}
   and  in Fig. \ref{fig:gcoex}.  
   As shown, $\gamma_{m}$ barely
    changes along the coexistence line when $p$ varies from strongly negative to 
     moderately positive. Interestingly, $\gamma_{m}$ displays  a shallow minimum.
  Thus, as long as $\Delta \mu$ does not change significantly
  with $\Delta T$ at negative $p$, 
  one can explain why
    in Fig.
       \ref{fig:nucleation}, $N_{c}$, $\Delta G_{c}$, $\gamma$,
        and $J$ seem to be independent of $p$ against the supercooling when $p$ is
         negative or moderate. 
         In order to
         confirm this, we 
         evaluate the effect of $p$ on $\Delta \mu$ 
          as a function of supercooling $\Delta T$ by 
          comparing with the value at 1 bar.
         To do so, we compute   $(\Delta \mu_{p} -\Delta \mu_{1} )/\Delta \mu_{1}$
           for the different isobars $p = $ -2600, -2000, -1000, 1, 2000 bar as a function of $\Delta T$ (for 1 and 2000 bar we use the data from Ref. \cite{espinosaPRL2016}). 
             As can be seen in Fig. \ref{fig:gcoex2} a),
              the 2000 bar isobar is very similar to the -1000 bar one
               in terms of $\Delta \mu$ with respect to $\Delta \mu_{1}$,
               and the -2600 bar is the one that deviates the most
                with up to 18$\%$. This deviation is however
                 compensated in $\gamma$ which is
                  rather dispersed and in the end $N_{c}$,
                   $\Delta G_{c}$, and $J$ are very well 
                   described by universal empirical expressions.
           Moreover, in Fig. \ref{fig:gcoex2} b),
           we show $\Delta G$ obtained as $N_{c}|\Delta \mu|/2$
             by setting 
           $N_c$ to the common fit of 
           Eq. \ref{eq:Nc} and changing $\Delta \mu$ to that
            of the different isobars. 
            As can be seen,
            from strongly
            negative to moderately positive
             pressure, the change in $\Delta \mu$ does not 
              significantly affect the free energy barrier for
               isobars between -2600 to 450 bar. 
           Thus, we confirm that the universality in nucleation properties  presented
              in Fig. \ref{fig:nucleation1} and Fig. \ref{fig:nucleation} is the consequence of the 
              small variation with
               $p$ of the difference in chemical
               potential $\Delta \mu$ as well as in 
               the interfacial free energy both at coexistence
               $\gamma_{m}$ and for the nucleation $\gamma$ at a given $\Delta T$. 
              \\
  
  \begin{table}
\begin{ruledtabular}
\begin{tabular}{c|c|c}
$p_m$ [bar]& $T_m$ [K] & $\gamma_{m}$ [mJ/m$^{2}$]  \\
\hline
 -2600 & 279.0  &         27.1(1.5)                     \\
  -2000 &  280.0 &            26.5(1.5)              \\
 -1000 & 278.0  &                 25.6(1.5)             \\
 1 & 270.0 & 27.2 (0.8)  \\
  1000 &  260.0 &                 29.0(1.5)           \\
2000 & 246.5  &           37.2(1.5)              \\
\end{tabular}
\end{ruledtabular}
\caption{ Interfacial free energy $\gamma_{m}$ at different $T-p$ points of the
 coexistence line for the basal plane.
 Value at 1 bar is from   Ref. \cite{espinosa2016ice}. }
\label{tab:gammas}
\end{table}

\subsection{Interfacial free energy and  melting line of the ice Ih-water interface}
       
         We now understand  the small variability with pressure 
    of the nucleation properties
   as a function of  supercooling at negative and
     moderate pressure. 
   In order to understand  why $\gamma_{m}$ displays a shallow minimum, 
   we use the  thermodynamic formalism of Gibbs for interfaces \cite{gibbs1928collected,rowlinson2013molecular}. 
      The interfacial Gibbs-Duhem relation is given by,

       \begin{equation}
    d\gamma_{m} =    - \Gamma d\mu_{m} - \eta_{\gamma}dT_{m},
    \label{eq:GDI}
    \end{equation}
      
   \noindent     where $\Gamma = N_{\gamma}/A$ is the surface excess density, also called adsorption,
    and $\eta_{\gamma} = S_{\gamma}/A$ is the excess contribution to the entropy. Since the location of the 
         dividing surface is arbitrary, excess functions depend also on this choice
          with the exception of $\gamma_{m}$.  
        For a planar interface, $\gamma_{m}$ does not
       change with the location of the dividing surface
        unlike in the case of  curved interfaces,  where $\gamma$ does change with its location \cite{rowlinson2013molecular,montero2020young,montero2022thermodynamics}.
            The choice that most simplifies the thermodynamic
             treatment  in our case is the equimolar dividing surface, usually denoted as the Gibbs dividing surface, 
             where the  excess components $N_{\gamma}$ is zero, and so is  $\Gamma$
             (see Appendix  for a general dividing surface treatment). 
             Hence, we can write 
     
     \begin{equation}
         \frac{d\gamma_{m}}{dT_{m}} = - \eta_{\gamma}^{e},
         \label{eq:adp}
     \end{equation}
      
      \noindent  where the superscript $e$ denotes the equimolar dividing surface. 
      Equation \ref{eq:adp} provides us with the temperature
       dependence of the interfacial free energy. 
       It is crucial to note that this  derivative must 
        be taken along the coexistence line so that $p$
         is not constant. In fact, we can change Eq. \ref{eq:adp}
          to describe the change of $\gamma_{m}$ with  pressure along the melting line $p_{m}$ as,
          
                 \begin{equation}
         \frac{d\gamma_{m}}{dp_{m}} = - \eta_{\gamma}^{e}\frac{dT_{m}}{dp_{m}} .
         \label{eq:imprt}
         \end{equation}

      In our case, Eq. \ref{eq:imprt} is 
      more convenient due to the reentrant behavior of the melting curve, 
      i.e. for each $T_m$ one has two values of  $p_m$ 
      whereas for each $p_m$ there is only one value of $T_m$ (see solid red curve in Fig. 3).
       From Eq. \ref{eq:imprt}, one can see that the change in $\gamma_{m}$ with
  $p_{m}$ is determined by the slope of the melting line
   and the value of the excess entropy per area at the
    equimolar dividing surface, $\eta_{\gamma}^{e}$.
   This means that
    if there is  reentrant behavior for the melting point, there must be
      reentrant  behavior also for $\gamma_{m}$ as a function of pressure exactly
      at the same $p_{m}$, because $\eta_{\gamma}^{e}$ must be finite. 
      In fact, Bianco et al. \cite{bianco2021anomalous} reported
       reentrant behavior in the ice Ih-liquid coexistence
       line of TIP4P/Ice, whose turning point occurred at -2000 bar.\\
        
\begin{figure}[h!]
\centering
\includegraphics[width=3.3in]{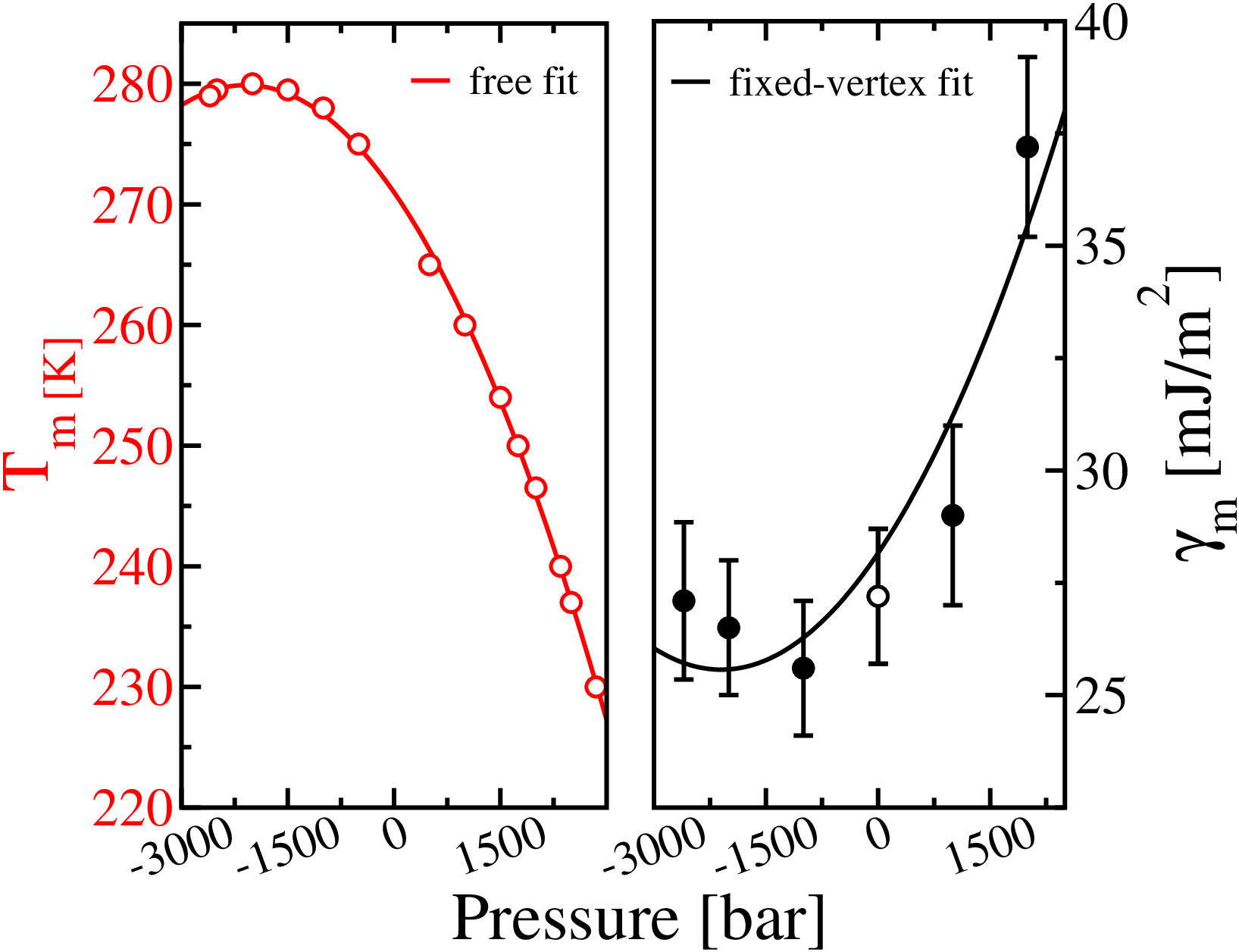}

\caption{\label{fig:2panel} Left: Melting temperature as a function of pressure. Empty circles are from Ref. \cite{bianco2021anomalous}. The line is a quadratic fit. Right: Interfacial free energy as a function of
 pressure. Solid points are from this work and empty points
  are from Ref. \cite{espinosa2016ice}. The line is a quadratic
   fit constrained to have the vertex at the same pressure (-2000 bar) than
    the quadratic fit of the left panel. }
\end{figure}

        Next, we want to confirm that the maximum in
         the melting line $T_{m}(p_{m})$ is consistent with the
          minimum in $\gamma_{m}(p_{m})$ that we have
           obtained from the mold integration technique. Thus,
          we fit the data for $\gamma_{m}(p_{m})$
            from mold integration with a quadratic fit
             with the constraint of having the vertex at the same
              $p_{m}$ (-2000 bar) as the quadratic
              fit for $T_{m}(p_{m})$. The latter, 
              $T_{m}(p) = a_{T_{m}}p^{2} + b_{T_{m}}p + c_{T_{m}}$
                 has the parameters $c_{T_{m}} = $271 K,  $b_{T_{m}} =-8.5\cdot 10^{-3}$ K/bar, and  $a_{T_{m}} =  - 2\cdot 10^{-6}$ K/bar$^{2}$.
              In this way, we
               assume that $\eta_{\gamma}^{e}$ is constant.
               The result  is shown
               in Fig. \ref{fig:2panel}. In the left  panel, we show  
                the melting line with points from the direct
                 coexistence simulations of Ref. \cite{bianco2021anomalous}
                  and the quadratic fit. On the right  panel,
                   we show the points of $\gamma_{m}$ from mold integration
                    from this work and Ref. \cite{espinosa2016ice}
                    as well as the quadratic fit. 
                    As can be seen in the right panel,  the fit is
                     fairly good even though we impose constant  $\eta_{\gamma}^{e}$ and quadratic fits with the 
                      constraint of having the vertex at the same $p$.
                      Therefore, assuming that $\eta_{\gamma}^{e}$
                       is constant seems to be a reasonable
                        approximation.\\
                        
                        At this level of approximation, $\eta_{\gamma}^{e}$
                        is found to be 0.32 mJ$/$m$^{2}$K.  Notice that
        $\eta_{\gamma}^{e} >0$ as expected from Eq. \ref{eq:imprt}.
        For instance, from 1 bar to 2000 bar, $T_{m}$ decreases from 270 K to 246.5 K,
         and $\gamma_{m}$ increases from 27.2 mJ/m$^2$ to  37.2  mJ/m$^2$. 
      Therefore, $d\gamma_{m}/dp_{m} > $ 0  and $dT_{m}/dp_{m} < $ 0, which
       means that $\eta_{\gamma}^{e}$ is positive. On the other side
        of the vertex, from -2600 bar to -2000 bar, $T_{m}$ increases from 279 K to 280K while
         $\gamma_{m}$ decreases from 27.1 mJ/m$^2$ to 26.5 mJ/m$^2$. Hence, $d\gamma_{m}/dp_{m} < $ 0  and $dT_{m}/dp_{m} > $ 0
         so that the same sign in $\eta_{\gamma}^{e}$ holds.
         Notice that Eq. \ref{eq:adp} 
         is only valid for planar interfaces along the melting line.
         If one tries to apply this equation away from of this line 
         as was done in previous works \cite{espinosaPRL2016,qiu2018water,piaggi2022}, 
         probably one should incorporate terms that account for the change in 
         $\gamma$ due to curvature.
         Notice also that the empirical relation proposed by Turnbull which
         states that $\gamma_{m}$ is
          proportional to the change in melting enthalpy $\Delta H_{m}$ does not
           describe $\gamma_{m}$ well at high pressure. 
          From 1 to 2000 bar,
          $\Delta H_{m}$ decreases from 1.44 kcal/mol 
          to almost 1
           kcal/mol in experiments \cite{eisenberg2005structure} and
           from 1.29 kcal/mol 
            to approximately 1  kcal/mol \cite{abascal2009triple}
            for the TIP4P/Ice model. 
            Thus, the Turnbull relation predicts a
              decreasing $\gamma_{m}$, which is not supported by our direct
              calculations via the mold integration technique.\\
         
         As can be seen in Fig. \ref{fig:2panel}, the knowledge of the equilibrium melting curve, and the assumption of a constant value for the interfacial excess entropy is sufficient  to understand the complex variation of $\gamma_m$ along the melting line. 
        Another relevant excess variable which depends
         on $\gamma_m$, $T_m$, and $\eta_{\gamma}^{e}$
         is the excess energy $e_{\gamma}^{e}$,
              
                 \begin{equation}
              e_{\gamma}^{e} = \gamma_{m} + T_{m}\eta_{\gamma}^{e}.
              \label{eq:exceses}
          \end{equation}

      The excess energy $e_{\gamma}^{e}$ is 
        the difference in energy between
                          the actual system having
                           an interface
                           and a virtual system where the two phases
                                     remain unchanged up to the dividing surface (the  equimolar one in this case).
        As a result of Eq. \ref{eq:imprt}, the following relation holds,

                         \begin{equation}
         \frac{de_{\gamma}^{e}}{dp_{m}} = T_{m}\frac{d \eta_{\gamma}^{e}}{dp_{m}} ,
         \label{eq:imprt22}
         \end{equation}
         
   \noindent        so that if $\eta_{\gamma}^{e}$ is constant, then $e_{\gamma}^{e}$ must
          be constant as well. If we approximate $\eta_{\gamma}^{e}$ as constant with the value of 0.32 mJ/m$^{2}$K,
            we find $e_{\gamma}^{e} = $ 115 mJ/m$^2$. \\

\section{Conclusions}

In conclusion, we perform seeding simulations to study  ice
 nucleation at negative pressures. Such conditions can be relevant in porous media and water transport
  in plants, where supercooled water can be at negative pressure. By comparing with previous results, we show that  universal empirical
 expressions describe 
$N_{c}$, $\Delta G_{c}$, $\gamma$, and $J$, as a function of
 supercooling for isobars in the regime from strongly 
 negative (-2600 bar) 
  to moderately positive pressures (500 bar).
  Only when
  pressure is high (2000 bar), these relations break down.
  In the regime where pressure hardly plays any role, 
  the isobaric Tolman length is predicted to be positive and roughly constant with the value of 0.24 nm. 
  Also, our results suggest that the 
  homogeneous nucleation line
   should be parallel to the coexistence line when
    pressure is  below approximately 500 bar (while at higher pressure they are not).
    We explain this result by inspecting  how  
   the interfacial free energy at coexistence changes with 
    pressure. We evaluate   the interfacial free energy at coexistence 
    at different states from strongly negative to high pressure 
    by means of the mold
     integration technique. We show that the interfacial free energy at coexistence
      barely changes with pressure as long as the system
       is below 500 bar.
In fact, a shallow minimum is reported at negative pressure suggesting that the minimum interfacial free
               energy between ice Ih and water is around 26 $\pm$ 1 mJ/m$^2$ for the basal
                plane 
               expanding for a broad
                 range of pressure centered around -2000 bar. 
    Then, we use the Gibbsian formalism to explain
     that this minimum in the interfacial free energy
     is connected to a maximum in the melting temperature as a function of pressure. 
     In particular, we show that the change
       in the interfacial free energy with pressure is 
       proportional to the
        excess entropy and to the slope of the melting line. 
        Thus, the reentrance in the interfacial free energy
         occurs because of the reentrance in the melting line,
         which happens due to the cross-over in density 
          between ice and water.
        Finally, we estimate the excess entropy and
         the excess energy of the ice Ih-water interface. 
         We suggest that a constant value of
          0.32mJ/m$^2$K and 115 mJ/m$^2$ respectively 
          is enough to provide a good description of the
           thermodynamics of the ice Ih-water interface.

\section{Acknowledgments}

 The authors thank Eduardo Sanz and Salvatore Romano for fruitful discussions. 
 PMdH ackowledges support from the SFB TACO (project nr. F81-N) funded by the
  Austrian Science Fund.  JRE acknowledges funding from the Oppenheimer Fellowship, the Roger Ekins Fellowship from Emmanuel College, and a Ramon y Cajal Fellowship (RYC2021‐030937‐I). CV acknowledges support from project  PID2019-105898GB-C21
of the Ministerio de Educacion y Cultura. This work has been performed using resources provided by the Spanish Supercomputing Network (RES), the Vienna Scientific Cluster (VSC), and the Cambridge Tier-2 system operated by the University of Cambridge Research Computing Service (http://www.hpc.cam.ac.uk) funded by EPSRC Tier-2 capital grant EP/P020259/1.

\section{Author declarations}

\subsection{Conflict of Interest}

The authors have no conflicts to disclose. 

\subsection{Data availability}

The data that support the findings of this study are available
from the corresponding author upon reasonable request.

\section{ Appendix: Interfacial free energy along the melting line for a general dividing surface}

In this work we used the equimolar dividing surface for simplicity. However, Eqs. \ref{eq:adp} and 
\ref{eq:imprt} can be generalized for any choice of the dividing surface. To do so, it is necessary
 to involve not only the interfacial Gibbs-Duhem relation (\ref{eq:GDI}), but also the
  ice and liquid Gibbs-Duhem relations. Respectively, these are,

         \begin{equation}
    d\mu_{m} - v_{i}dp_{m} + s_{i} dT_{m} = 0,
    \label{eq:GDice}
    \end{equation}
    
             \begin{equation}
    d\mu_{m} - v_{w}dp_{m} + s_{w} dT_{m} = 0,
      \label{eq:GDwat}
    \end{equation}
    
    \noindent  where $v$ is the volume per molecule (the inverse of the number density)
     and $s$ is the entropy per molecule. Since phase equilibrium holds, $d\mu_{m}$, $dp_{m}$, and $dT_{m}$ are common in all phases. Notice that
      from Eq. \ref{eq:GDice} and Eq. \ref{eq:GDwat}, one can obtain the Clausius-Clapeyron
       relation that explains the slope of the melting line. 
       
       \begin{equation}
           \frac{dT_{m}}{dp_{m}} = \frac{v_{w} - v_{i}}{s_{w} - s_{i}}.
       \end{equation}
     
     By including also Eq. \ref{eq:GDI} in the relation, one can obtain the temperature  and pressure
       dependence of the interfacial free energy without imposing a specific dividing surface. 
       For the temperature, one obtains,
       
           \begin{equation}
           \frac{d\gamma_{m}}{dT_{m}} = \left[ \Gamma  \frac{ v_{w}s_{i} - v_{i}s_{w}}{v_{w} - v_{i}}     -\eta_{\gamma}     \right] ,
       \end{equation}
       
  \noindent    whereas for the pressure, one finds
    
           \begin{equation}
           \frac{d\gamma_{m}}{dp_{m}} = \left[ \Gamma  \frac{ v_{w}s_{i} - v_{i}s_{w}}{v_{w} - v_{i}}     -\eta_{\gamma}     \right] \frac{dT_{m}}{dp_{m}}.
       \end{equation}
    
    As can be seen, at the equimolar dividing surface where $\Gamma = 0$, one recovers
     Eq. \ref{eq:adp} and Eq. \ref{eq:imprt} respectively. These expressions are relevant
      when nucleation data are extrapolated to coexistence because the relevant dividing surface
     in nucleation  is usually the surface of tension for which $\Gamma \neq 0$.
    

\bibliography{nuevabib}

\begin{thebibliography}{67}%
\makeatletter
\providecommand \@ifxundefined [1]{%
 \@ifx{#1\undefined}
}%
\providecommand \@ifnum [1]{%
 \ifnum #1\expandafter \@firstoftwo
 \else \expandafter \@secondoftwo
 \fi
}%
\providecommand \@ifx [1]{%
 \ifx #1\expandafter \@firstoftwo
 \else \expandafter \@secondoftwo
 \fi
}%
\providecommand \natexlab [1]{#1}%
\providecommand \enquote  [1]{``#1''}%
\providecommand \bibnamefont  [1]{#1}%
\providecommand \bibfnamefont [1]{#1}%
\providecommand \citenamefont [1]{#1}%
\providecommand \href@noop [0]{\@secondoftwo}%
\providecommand \href [0]{\begingroup \@sanitize@url \@href}%
\providecommand \@href[1]{\@@startlink{#1}\@@href}%
\providecommand \@@href[1]{\endgroup#1\@@endlink}%
\providecommand \@sanitize@url [0]{\catcode `\\12\catcode `\$12\catcode
  `\&12\catcode `\#12\catcode `\^12\catcode `\_12\catcode `\%12\relax}%
\providecommand \@@startlink[1]{}%
\providecommand \@@endlink[0]{}%
\providecommand \url  [0]{\begingroup\@sanitize@url \@url }%
\providecommand \@url [1]{\endgroup\@href {#1}{\urlprefix }}%
\providecommand \urlprefix  [0]{URL }%
\providecommand \Eprint [0]{\href }%
\providecommand \doibase [0]{http://dx.doi.org/}%
\providecommand \selectlanguage [0]{\@gobble}%
\providecommand \bibinfo  [0]{\@secondoftwo}%
\providecommand \bibfield  [0]{\@secondoftwo}%
\providecommand \translation [1]{[#1]}%
\providecommand \BibitemOpen [0]{}%
\providecommand \bibitemStop [0]{}%
\providecommand \bibitemNoStop [0]{.\EOS\space}%
\providecommand \EOS [0]{\spacefactor3000\relax}%
\providecommand \BibitemShut  [1]{\csname bibitem#1\endcsname}%
\let\auto@bib@innerbib\@empty
\bibitem [{\citenamefont {Geidobler}\ and\ \citenamefont
  {Winter}(2013)}]{geidobler2013controlled}%
  \BibitemOpen
  \bibfield  {author} {\bibinfo {author} {\bibfnamefont {R.}~\bibnamefont
  {Geidobler}}\ and\ \bibinfo {author} {\bibfnamefont {G.}~\bibnamefont
  {Winter}},\ }\href@noop {} {\bibfield  {journal} {\bibinfo  {journal}
  {European Journal of Pharmaceutics and Biopharmaceutics}\ }\textbf {\bibinfo
  {volume} {85}},\ \bibinfo {pages} {214} (\bibinfo {year} {2013})}\BibitemShut
  {NoStop}%
\bibitem [{\citenamefont {Xue}\ \emph {et~al.}(2015)\citenamefont {Xue},
  \citenamefont {Jin}, \citenamefont {He},\ and\ \citenamefont
  {Liu}}]{xue2015quantifying}%
  \BibitemOpen
  \bibfield  {author} {\bibinfo {author} {\bibfnamefont {X.}~\bibnamefont
  {Xue}}, \bibinfo {author} {\bibfnamefont {H.-L.}\ \bibnamefont {Jin}},
  \bibinfo {author} {\bibfnamefont {Z.-Z.}\ \bibnamefont {He}}, \ and\ \bibinfo
  {author} {\bibfnamefont {J.}~\bibnamefont {Liu}},\ }\href@noop {} {\bibfield
  {journal} {\bibinfo  {journal} {Journal of Heat Transfer}\ }\textbf {\bibinfo
  {volume} {137}},\ \bibinfo {pages} {091020} (\bibinfo {year}
  {2015})}\BibitemShut {NoStop}%
\bibitem [{\citenamefont {Pegg}(2009)}]{pegg2009principles}%
  \BibitemOpen
  \bibfield  {author} {\bibinfo {author} {\bibfnamefont {D.~E.}\ \bibnamefont
  {Pegg}},\ }\href@noop {} {\bibfield  {journal} {\bibinfo  {journal}
  {Preservation of Human oocytes CRC Press}\ } (\bibinfo {year}
  {2009})}\BibitemShut {NoStop}%
\bibitem [{\citenamefont {Kanno}\ \emph {et~al.}(1975)\citenamefont {Kanno},
  \citenamefont {Speedy},\ and\ \citenamefont
  {Angell}}]{kanno1975supercooling}%
  \BibitemOpen
  \bibfield  {author} {\bibinfo {author} {\bibfnamefont {H.}~\bibnamefont
  {Kanno}}, \bibinfo {author} {\bibfnamefont {R.}~\bibnamefont {Speedy}}, \
  and\ \bibinfo {author} {\bibfnamefont {C.}~\bibnamefont {Angell}},\
  }\href@noop {} {\bibfield  {journal} {\bibinfo  {journal} {Science}\ }\textbf
  {\bibinfo {volume} {189}},\ \bibinfo {pages} {880} (\bibinfo {year}
  {1975})}\BibitemShut {NoStop}%
\bibitem [{\citenamefont {Kalichevsky}\ \emph {et~al.}(1995)\citenamefont
  {Kalichevsky}, \citenamefont {Knorr},\ and\ \citenamefont
  {Lillford}}]{kalichevsky1995potential}%
  \BibitemOpen
  \bibfield  {author} {\bibinfo {author} {\bibfnamefont {M.}~\bibnamefont
  {Kalichevsky}}, \bibinfo {author} {\bibfnamefont {D.}~\bibnamefont {Knorr}},
  \ and\ \bibinfo {author} {\bibfnamefont {P.}~\bibnamefont {Lillford}},\
  }\href@noop {} {\bibfield  {journal} {\bibinfo  {journal} {Trends in Food
  Science \& Technology}\ }\textbf {\bibinfo {volume} {6}},\ \bibinfo {pages}
  {253} (\bibinfo {year} {1995})}\BibitemShut {NoStop}%
\bibitem [{\citenamefont {Martino}\ \emph {et~al.}(1998)\citenamefont
  {Martino}, \citenamefont {Otero}, \citenamefont {Sanz},\ and\ \citenamefont
  {Zaritzky}}]{martino1998size}%
  \BibitemOpen
  \bibfield  {author} {\bibinfo {author} {\bibfnamefont {M.~N.}\ \bibnamefont
  {Martino}}, \bibinfo {author} {\bibfnamefont {L.}~\bibnamefont {Otero}},
  \bibinfo {author} {\bibfnamefont {P.}~\bibnamefont {Sanz}}, \ and\ \bibinfo
  {author} {\bibfnamefont {N.}~\bibnamefont {Zaritzky}},\ }\href@noop {}
  {\bibfield  {journal} {\bibinfo  {journal} {Meat Science}\ }\textbf {\bibinfo
  {volume} {50}},\ \bibinfo {pages} {303} (\bibinfo {year} {1998})}\BibitemShut
  {NoStop}%
\bibitem [{\citenamefont {Espinosa}\ \emph
  {et~al.}(2016{\natexlab{a}})\citenamefont {Espinosa}, \citenamefont
  {Zaragoza}, \citenamefont {Rosales-Pelaez}, \citenamefont {Navarro},
  \citenamefont {Valeriani}, \citenamefont {Vega},\ and\ \citenamefont
  {Sanz}}]{espinosaPRL2016}%
  \BibitemOpen
  \bibfield  {author} {\bibinfo {author} {\bibfnamefont {J.~R.}\ \bibnamefont
  {Espinosa}}, \bibinfo {author} {\bibfnamefont {A.}~\bibnamefont {Zaragoza}},
  \bibinfo {author} {\bibfnamefont {P.}~\bibnamefont {Rosales-Pelaez}},
  \bibinfo {author} {\bibfnamefont {C.}~\bibnamefont {Navarro}}, \bibinfo
  {author} {\bibfnamefont {C.}~\bibnamefont {Valeriani}}, \bibinfo {author}
  {\bibfnamefont {C.}~\bibnamefont {Vega}}, \ and\ \bibinfo {author}
  {\bibfnamefont {E.}~\bibnamefont {Sanz}},\ }\href@noop {} {\bibfield
  {journal} {\bibinfo  {journal} {Physical Review Letters}\ }\textbf {\bibinfo
  {volume} {117}},\ \bibinfo {pages} {135702} (\bibinfo {year}
  {2016}{\natexlab{a}})}\BibitemShut {NoStop}%
\bibitem [{\citenamefont {Tolman}(1949)}]{tolman1949effect}%
  \BibitemOpen
  \bibfield  {author} {\bibinfo {author} {\bibfnamefont {R.~C.}\ \bibnamefont
  {Tolman}},\ }\href@noop {} {\bibfield  {journal} {\bibinfo  {journal} {The
  journal of Chemical Physics}\ }\textbf {\bibinfo {volume} {17}},\ \bibinfo
  {pages} {333} (\bibinfo {year} {1949})}\BibitemShut {NoStop}%
\bibitem [{\citenamefont {Sanz}\ \emph {et~al.}(2013)\citenamefont {Sanz},
  \citenamefont {Vega}, \citenamefont {Espinosa}, \citenamefont
  {Caballero-Bernal}, \citenamefont {Abascal},\ and\ \citenamefont
  {Valeriani}}]{sanz2013homogeneous}%
  \BibitemOpen
  \bibfield  {author} {\bibinfo {author} {\bibfnamefont {E.}~\bibnamefont
  {Sanz}}, \bibinfo {author} {\bibfnamefont {C.}~\bibnamefont {Vega}}, \bibinfo
  {author} {\bibfnamefont {J.}~\bibnamefont {Espinosa}}, \bibinfo {author}
  {\bibfnamefont {R.}~\bibnamefont {Caballero-Bernal}}, \bibinfo {author}
  {\bibfnamefont {J.}~\bibnamefont {Abascal}}, \ and\ \bibinfo {author}
  {\bibfnamefont {C.}~\bibnamefont {Valeriani}},\ }\href@noop {} {\bibfield
  {journal} {\bibinfo  {journal} {Journal of the American Chemical Society}\
  }\textbf {\bibinfo {volume} {135}},\ \bibinfo {pages} {15008} (\bibinfo
  {year} {2013})}\BibitemShut {NoStop}%
\bibitem [{\citenamefont {Koop}\ \emph {et~al.}(2000)\citenamefont {Koop},
  \citenamefont {Luo}, \citenamefont {Tsias},\ and\ \citenamefont
  {Peter}}]{koop2000water}%
  \BibitemOpen
  \bibfield  {author} {\bibinfo {author} {\bibfnamefont {T.}~\bibnamefont
  {Koop}}, \bibinfo {author} {\bibfnamefont {B.}~\bibnamefont {Luo}}, \bibinfo
  {author} {\bibfnamefont {A.}~\bibnamefont {Tsias}}, \ and\ \bibinfo {author}
  {\bibfnamefont {T.}~\bibnamefont {Peter}},\ }\href@noop {} {\bibfield
  {journal} {\bibinfo  {journal} {Nature}\ }\textbf {\bibinfo {volume} {406}},\
  \bibinfo {pages} {611} (\bibinfo {year} {2000})}\BibitemShut {NoStop}%
\bibitem [{\citenamefont {Espinosa}\ \emph {et~al.}(2014)\citenamefont
  {Espinosa}, \citenamefont {Sanz}, \citenamefont {Valeriani},\ and\
  \citenamefont {Vega}}]{espinosa2014homogeneous}%
  \BibitemOpen
  \bibfield  {author} {\bibinfo {author} {\bibfnamefont {J.}~\bibnamefont
  {Espinosa}}, \bibinfo {author} {\bibfnamefont {E.}~\bibnamefont {Sanz}},
  \bibinfo {author} {\bibfnamefont {C.}~\bibnamefont {Valeriani}}, \ and\
  \bibinfo {author} {\bibfnamefont {C.}~\bibnamefont {Vega}},\ }\href@noop {}
  {\bibfield  {journal} {\bibinfo  {journal} {The Journal of Chemical Physics}\
  }\textbf {\bibinfo {volume} {141}},\ \bibinfo {pages} {18C529} (\bibinfo
  {year} {2014})}\BibitemShut {NoStop}%
\bibitem [{\citenamefont {Niu}\ \emph {et~al.}(2019)\citenamefont {Niu},
  \citenamefont {Yang},\ and\ \citenamefont {Parrinello}}]{niu2019temperature}%
  \BibitemOpen
  \bibfield  {author} {\bibinfo {author} {\bibfnamefont {H.}~\bibnamefont
  {Niu}}, \bibinfo {author} {\bibfnamefont {Y.~I.}\ \bibnamefont {Yang}}, \
  and\ \bibinfo {author} {\bibfnamefont {M.}~\bibnamefont {Parrinello}},\
  }\href@noop {} {\bibfield  {journal} {\bibinfo  {journal} {Physical Review
  Letters}\ }\textbf {\bibinfo {volume} {122}},\ \bibinfo {pages} {245501}
  (\bibinfo {year} {2019})}\BibitemShut {NoStop}%
\bibitem [{\citenamefont {Espinosa}\ \emph
  {et~al.}(2016{\natexlab{b}})\citenamefont {Espinosa}, \citenamefont
  {Navarro}, \citenamefont {Sanz}, \citenamefont {Valeriani},\ and\
  \citenamefont {Vega}}]{espinosa2016time}%
  \BibitemOpen
  \bibfield  {author} {\bibinfo {author} {\bibfnamefont {J.}~\bibnamefont
  {Espinosa}}, \bibinfo {author} {\bibfnamefont {C.}~\bibnamefont {Navarro}},
  \bibinfo {author} {\bibfnamefont {E.}~\bibnamefont {Sanz}}, \bibinfo {author}
  {\bibfnamefont {C.}~\bibnamefont {Valeriani}}, \ and\ \bibinfo {author}
  {\bibfnamefont {C.}~\bibnamefont {Vega}},\ }\href@noop {} {\bibfield
  {journal} {\bibinfo  {journal} {The Journal of Chemical Physics}\ }\textbf
  {\bibinfo {volume} {145}},\ \bibinfo {pages} {211922} (\bibinfo {year}
  {2016}{\natexlab{b}})}\BibitemShut {NoStop}%
\bibitem [{\citenamefont {Piaggi}\ \emph {et~al.}(2022)\citenamefont {Piaggi},
  \citenamefont {Weis}, \citenamefont {Panagiotopoulos}, \citenamefont
  {Debenedetti},\ and\ \citenamefont {Car}}]{piaggi2022}%
  \BibitemOpen
  \bibfield  {author} {\bibinfo {author} {\bibfnamefont {P.~M.}\ \bibnamefont
  {Piaggi}}, \bibinfo {author} {\bibfnamefont {J.}~\bibnamefont {Weis}},
  \bibinfo {author} {\bibfnamefont {A.~Z.}\ \bibnamefont {Panagiotopoulos}},
  \bibinfo {author} {\bibfnamefont {P.~G.}\ \bibnamefont {Debenedetti}}, \ and\
  \bibinfo {author} {\bibfnamefont {R.}~\bibnamefont {Car}},\ }\href@noop {}
  {\bibfield  {journal} {\bibinfo  {journal} {Proceedings of the National
  Academy of Sciences}\ }\textbf {\bibinfo {volume} {119}} (\bibinfo {year}
  {2022})}\BibitemShut {NoStop}%
\bibitem [{\citenamefont {Li}\ \emph {et~al.}(2011)\citenamefont {Li},
  \citenamefont {Donadio}, \citenamefont {Russo},\ and\ \citenamefont
  {Galli}}]{li2011homogeneous}%
  \BibitemOpen
  \bibfield  {author} {\bibinfo {author} {\bibfnamefont {T.}~\bibnamefont
  {Li}}, \bibinfo {author} {\bibfnamefont {D.}~\bibnamefont {Donadio}},
  \bibinfo {author} {\bibfnamefont {G.}~\bibnamefont {Russo}}, \ and\ \bibinfo
  {author} {\bibfnamefont {G.}~\bibnamefont {Galli}},\ }\href@noop {}
  {\bibfield  {journal} {\bibinfo  {journal} {Physical Chemistry Chemical
  Physics}\ }\textbf {\bibinfo {volume} {13}},\ \bibinfo {pages} {19807}
  (\bibinfo {year} {2011})}\BibitemShut {NoStop}%
\bibitem [{\citenamefont {Laksmono}\ \emph {et~al.}(2015)\citenamefont
  {Laksmono}, \citenamefont {McQueen}, \citenamefont {Sellberg}, \citenamefont
  {Loh}, \citenamefont {Huang}, \citenamefont {Schlesinger}, \citenamefont
  {Sierra}, \citenamefont {Hampton}, \citenamefont {Nordlund}, \citenamefont
  {Beye} \emph {et~al.}}]{laksmono2015anomalous}%
  \BibitemOpen
  \bibfield  {author} {\bibinfo {author} {\bibfnamefont {H.}~\bibnamefont
  {Laksmono}}, \bibinfo {author} {\bibfnamefont {T.~A.}\ \bibnamefont
  {McQueen}}, \bibinfo {author} {\bibfnamefont {J.~A.}\ \bibnamefont
  {Sellberg}}, \bibinfo {author} {\bibfnamefont {N.~D.}\ \bibnamefont {Loh}},
  \bibinfo {author} {\bibfnamefont {C.}~\bibnamefont {Huang}}, \bibinfo
  {author} {\bibfnamefont {D.}~\bibnamefont {Schlesinger}}, \bibinfo {author}
  {\bibfnamefont {R.~G.}\ \bibnamefont {Sierra}}, \bibinfo {author}
  {\bibfnamefont {C.~Y.}\ \bibnamefont {Hampton}}, \bibinfo {author}
  {\bibfnamefont {D.}~\bibnamefont {Nordlund}}, \bibinfo {author}
  {\bibfnamefont {M.}~\bibnamefont {Beye}},  \emph {et~al.},\ }\href@noop {}
  {\bibfield  {journal} {\bibinfo  {journal} {The Journal of Physical Chemistry
  Letters}\ }\textbf {\bibinfo {volume} {6}},\ \bibinfo {pages} {2826}
  (\bibinfo {year} {2015})}\BibitemShut {NoStop}%
\bibitem [{\citenamefont {Amaya}\ and\ \citenamefont
  {Wyslouzil}(2018)}]{amaya2018ice}%
  \BibitemOpen
  \bibfield  {author} {\bibinfo {author} {\bibfnamefont {A.~J.}\ \bibnamefont
  {Amaya}}\ and\ \bibinfo {author} {\bibfnamefont {B.~E.}\ \bibnamefont
  {Wyslouzil}},\ }\href@noop {} {\bibfield  {journal} {\bibinfo  {journal} {The
  Journal of Chemical Physics}\ }\textbf {\bibinfo {volume} {148}},\ \bibinfo
  {pages} {084501} (\bibinfo {year} {2018})}\BibitemShut {NoStop}%
\bibitem [{\citenamefont {Jeffery}\ and\ \citenamefont
  {Austin}(1997)}]{jeffery1997homogeneous}%
  \BibitemOpen
  \bibfield  {author} {\bibinfo {author} {\bibfnamefont {C.}~\bibnamefont
  {Jeffery}}\ and\ \bibinfo {author} {\bibfnamefont {P.}~\bibnamefont
  {Austin}},\ }\href@noop {} {\bibfield  {journal} {\bibinfo  {journal}
  {Journal of Geophysical Research: Atmospheres}\ }\textbf {\bibinfo {volume}
  {102}},\ \bibinfo {pages} {25269} (\bibinfo {year} {1997})}\BibitemShut
  {NoStop}%
\bibitem [{\citenamefont {Cheng}\ \emph {et~al.}(2018)\citenamefont {Cheng},
  \citenamefont {Dellago},\ and\ \citenamefont
  {Ceriotti}}]{cheng2018theoretical}%
  \BibitemOpen
  \bibfield  {author} {\bibinfo {author} {\bibfnamefont {B.}~\bibnamefont
  {Cheng}}, \bibinfo {author} {\bibfnamefont {C.}~\bibnamefont {Dellago}}, \
  and\ \bibinfo {author} {\bibfnamefont {M.}~\bibnamefont {Ceriotti}},\
  }\href@noop {} {\bibfield  {journal} {\bibinfo  {journal} {Physical Chemistry
  Chemical Physics}\ }\textbf {\bibinfo {volume} {20}},\ \bibinfo {pages}
  {28732} (\bibinfo {year} {2018})}\BibitemShut {NoStop}%
\bibitem [{\citenamefont {Marcolli}(2017)}]{marcolli2017ice}%
  \BibitemOpen
  \bibfield  {author} {\bibinfo {author} {\bibfnamefont {C.}~\bibnamefont
  {Marcolli}},\ }\href@noop {} {\bibfield  {journal} {\bibinfo  {journal}
  {Scientific reports}\ }\textbf {\bibinfo {volume} {7}},\ \bibinfo {pages} {1}
  (\bibinfo {year} {2017})}\BibitemShut {NoStop}%
\bibitem [{\citenamefont {Bianco}\ \emph {et~al.}(2021)\citenamefont {Bianco},
  \citenamefont {de~Hijes}, \citenamefont {Lamas}, \citenamefont {Sanz},\ and\
  \citenamefont {Vega}}]{bianco2021anomalous}%
  \BibitemOpen
  \bibfield  {author} {\bibinfo {author} {\bibfnamefont {V.}~\bibnamefont
  {Bianco}}, \bibinfo {author} {\bibfnamefont {P.~M.}\ \bibnamefont
  {de~Hijes}}, \bibinfo {author} {\bibfnamefont {C.~P.}\ \bibnamefont {Lamas}},
  \bibinfo {author} {\bibfnamefont {E.}~\bibnamefont {Sanz}}, \ and\ \bibinfo
  {author} {\bibfnamefont {C.}~\bibnamefont {Vega}},\ }\href@noop {} {\bibfield
   {journal} {\bibinfo  {journal} {Physical Review Letters}\ }\textbf {\bibinfo
  {volume} {126}},\ \bibinfo {pages} {015704} (\bibinfo {year}
  {2021})}\BibitemShut {NoStop}%
\bibitem [{\citenamefont {Rosky}\ \emph {et~al.}(2022)\citenamefont {Rosky},
  \citenamefont {Cantrell}, \citenamefont {Li},\ and\ \citenamefont
  {Shaw}}]{rosky2022homogeneous}%
  \BibitemOpen
  \bibfield  {author} {\bibinfo {author} {\bibfnamefont {E.}~\bibnamefont
  {Rosky}}, \bibinfo {author} {\bibfnamefont {W.}~\bibnamefont {Cantrell}},
  \bibinfo {author} {\bibfnamefont {T.}~\bibnamefont {Li}}, \ and\ \bibinfo
  {author} {\bibfnamefont {R.~A.}\ \bibnamefont {Shaw}},\ }\href@noop {}
  {\bibfield  {journal} {\bibinfo  {journal} {Chemical Physics Letters}\
  }\textbf {\bibinfo {volume} {789}},\ \bibinfo {pages} {139289} (\bibinfo
  {year} {2022})}\BibitemShut {NoStop}%
\bibitem [{\citenamefont {Roedder}(1967)}]{roedder1967metastable}%
  \BibitemOpen
  \bibfield  {author} {\bibinfo {author} {\bibfnamefont {E.}~\bibnamefont
  {Roedder}},\ }\href@noop {} {\bibfield  {journal} {\bibinfo  {journal}
  {Science}\ }\textbf {\bibinfo {volume} {155}},\ \bibinfo {pages} {1413}
  (\bibinfo {year} {1967})}\BibitemShut {NoStop}%
\bibitem [{\citenamefont {Wheeler}\ and\ \citenamefont
  {Stroock}(2008)}]{wheeler2008transpiration}%
  \BibitemOpen
  \bibfield  {author} {\bibinfo {author} {\bibfnamefont {T.~D.}\ \bibnamefont
  {Wheeler}}\ and\ \bibinfo {author} {\bibfnamefont {A.~D.}\ \bibnamefont
  {Stroock}},\ }\href@noop {} {\bibfield  {journal} {\bibinfo  {journal}
  {Nature}\ }\textbf {\bibinfo {volume} {455}},\ \bibinfo {pages} {208}
  (\bibinfo {year} {2008})}\BibitemShut {NoStop}%
\bibitem [{\citenamefont {Caupin}\ \emph {et~al.}(2012)\citenamefont {Caupin},
  \citenamefont {Arvengas}, \citenamefont {Davitt}, \citenamefont {Azouzi},
  \citenamefont {Shmulovich}, \citenamefont {Ramboz}, \citenamefont {Sessoms},\
  and\ \citenamefont {Stroock}}]{caupin2012exploring}%
  \BibitemOpen
  \bibfield  {author} {\bibinfo {author} {\bibfnamefont {F.}~\bibnamefont
  {Caupin}}, \bibinfo {author} {\bibfnamefont {A.}~\bibnamefont {Arvengas}},
  \bibinfo {author} {\bibfnamefont {K.}~\bibnamefont {Davitt}}, \bibinfo
  {author} {\bibfnamefont {M.~E.~M.}\ \bibnamefont {Azouzi}}, \bibinfo {author}
  {\bibfnamefont {K.~I.}\ \bibnamefont {Shmulovich}}, \bibinfo {author}
  {\bibfnamefont {C.}~\bibnamefont {Ramboz}}, \bibinfo {author} {\bibfnamefont
  {D.~A.}\ \bibnamefont {Sessoms}}, \ and\ \bibinfo {author} {\bibfnamefont
  {A.~D.}\ \bibnamefont {Stroock}},\ }\href@noop {} {\bibfield  {journal}
  {\bibinfo  {journal} {Journal of Physics: Condensed Matter}\ }\textbf
  {\bibinfo {volume} {24}},\ \bibinfo {pages} {284110} (\bibinfo {year}
  {2012})}\BibitemShut {NoStop}%
\bibitem [{\citenamefont {Debenedetti}(2021)}]{debenedetti2021metastable}%
  \BibitemOpen
  \bibfield  {author} {\bibinfo {author} {\bibfnamefont {P.~G.}\ \bibnamefont
  {Debenedetti}},\ }in\ \href@noop {} {\emph {\bibinfo {booktitle} {Metastable
  Liquids}}}\ (\bibinfo  {publisher} {Princeton university press},\ \bibinfo
  {year} {2021})\BibitemShut {NoStop}%
\bibitem [{\citenamefont {Imre}(2007)}]{imre2007existence}%
  \BibitemOpen
  \bibfield  {author} {\bibinfo {author} {\bibfnamefont {A.~R.}\ \bibnamefont
  {Imre}},\ }\href@noop {} {\bibfield  {journal} {\bibinfo  {journal} {Physica
  status solidi (b)}\ }\textbf {\bibinfo {volume} {244}},\ \bibinfo {pages}
  {893} (\bibinfo {year} {2007})}\BibitemShut {NoStop}%
\bibitem [{\citenamefont {Menzl}\ \emph {et~al.}(2016)\citenamefont {Menzl},
  \citenamefont {Gonzalez}, \citenamefont {Geiger}, \citenamefont {Caupin},
  \citenamefont {Abascal}, \citenamefont {Valeriani},\ and\ \citenamefont
  {Dellago}}]{menzl2016molecular}%
  \BibitemOpen
  \bibfield  {author} {\bibinfo {author} {\bibfnamefont {G.}~\bibnamefont
  {Menzl}}, \bibinfo {author} {\bibfnamefont {M.~A.}\ \bibnamefont {Gonzalez}},
  \bibinfo {author} {\bibfnamefont {P.}~\bibnamefont {Geiger}}, \bibinfo
  {author} {\bibfnamefont {F.}~\bibnamefont {Caupin}}, \bibinfo {author}
  {\bibfnamefont {J.~L.}\ \bibnamefont {Abascal}}, \bibinfo {author}
  {\bibfnamefont {C.}~\bibnamefont {Valeriani}}, \ and\ \bibinfo {author}
  {\bibfnamefont {C.}~\bibnamefont {Dellago}},\ }\href@noop {} {\bibfield
  {journal} {\bibinfo  {journal} {Proceedings of the National Academy of
  Sciences}\ }\textbf {\bibinfo {volume} {113}},\ \bibinfo {pages} {13582}
  (\bibinfo {year} {2016})}\BibitemShut {NoStop}%
\bibitem [{\citenamefont {Caupin}\ and\ \citenamefont
  {Herbert}(2006)}]{caupin2006cavitation}%
  \BibitemOpen
  \bibfield  {author} {\bibinfo {author} {\bibfnamefont {F.}~\bibnamefont
  {Caupin}}\ and\ \bibinfo {author} {\bibfnamefont {E.}~\bibnamefont
  {Herbert}},\ }\href@noop {} {\bibfield  {journal} {\bibinfo  {journal}
  {Comptes Rendus Physique}\ }\textbf {\bibinfo {volume} {7}},\ \bibinfo
  {pages} {1000} (\bibinfo {year} {2006})}\BibitemShut {NoStop}%
\bibitem [{\citenamefont {Henderson}\ and\ \citenamefont
  {Speedy}(1980)}]{henderson1980berthelot}%
  \BibitemOpen
  \bibfield  {author} {\bibinfo {author} {\bibfnamefont {S.}~\bibnamefont
  {Henderson}}\ and\ \bibinfo {author} {\bibfnamefont {R.}~\bibnamefont
  {Speedy}},\ }\href@noop {} {\bibfield  {journal} {\bibinfo  {journal}
  {Journal of Physics E: Scientific Instruments}\ }\textbf {\bibinfo {volume}
  {13}},\ \bibinfo {pages} {778} (\bibinfo {year} {1980})}\BibitemShut
  {NoStop}%
\bibitem [{\citenamefont {Briggs}(1950)}]{briggs1950limiting}%
  \BibitemOpen
  \bibfield  {author} {\bibinfo {author} {\bibfnamefont {L.~J.}\ \bibnamefont
  {Briggs}},\ }\href@noop {} {\bibfield  {journal} {\bibinfo  {journal}
  {Journal of Applied Physics}\ }\textbf {\bibinfo {volume} {21}},\ \bibinfo
  {pages} {721} (\bibinfo {year} {1950})}\BibitemShut {NoStop}%
\bibitem [{\citenamefont {Davitt}\ \emph {et~al.}(2010)\citenamefont {Davitt},
  \citenamefont {Rolley}, \citenamefont {Caupin}, \citenamefont {Arvengas},\
  and\ \citenamefont {Balibar}}]{davitt2010equation}%
  \BibitemOpen
  \bibfield  {author} {\bibinfo {author} {\bibfnamefont {K.}~\bibnamefont
  {Davitt}}, \bibinfo {author} {\bibfnamefont {E.}~\bibnamefont {Rolley}},
  \bibinfo {author} {\bibfnamefont {F.}~\bibnamefont {Caupin}}, \bibinfo
  {author} {\bibfnamefont {A.}~\bibnamefont {Arvengas}}, \ and\ \bibinfo
  {author} {\bibfnamefont {S.}~\bibnamefont {Balibar}},\ }\href@noop {}
  {\bibfield  {journal} {\bibinfo  {journal} {The Journal of Chemical Physics}\
  }\textbf {\bibinfo {volume} {133}},\ \bibinfo {pages} {174507} (\bibinfo
  {year} {2010})}\BibitemShut {NoStop}%
\bibitem [{\citenamefont {Abascal}\ \emph {et~al.}(2005)\citenamefont
  {Abascal}, \citenamefont {Sanz}, \citenamefont {Garc{\'\i}a~Fern{\'a}ndez},\
  and\ \citenamefont {Vega}}]{abascal2005potential}%
  \BibitemOpen
  \bibfield  {author} {\bibinfo {author} {\bibfnamefont {J.}~\bibnamefont
  {Abascal}}, \bibinfo {author} {\bibfnamefont {E.}~\bibnamefont {Sanz}},
  \bibinfo {author} {\bibfnamefont {R.}~\bibnamefont
  {Garc{\'\i}a~Fern{\'a}ndez}}, \ and\ \bibinfo {author} {\bibfnamefont
  {C.}~\bibnamefont {Vega}},\ }\href@noop {} {\bibfield  {journal} {\bibinfo
  {journal} {The Journal of Chemical Physics}\ }\textbf {\bibinfo {volume}
  {122}},\ \bibinfo {pages} {234511} (\bibinfo {year} {2005})}\BibitemShut
  {NoStop}%
\bibitem [{\citenamefont {Weiss}\ \emph {et~al.}(2011)\citenamefont {Weiss},
  \citenamefont {Rullich}, \citenamefont {K{\"o}hler},\ and\ \citenamefont
  {Frauenheim}}]{weiss2011kinetic}%
  \BibitemOpen
  \bibfield  {author} {\bibinfo {author} {\bibfnamefont {V.~C.}\ \bibnamefont
  {Weiss}}, \bibinfo {author} {\bibfnamefont {M.}~\bibnamefont {Rullich}},
  \bibinfo {author} {\bibfnamefont {C.}~\bibnamefont {K{\"o}hler}}, \ and\
  \bibinfo {author} {\bibfnamefont {T.}~\bibnamefont {Frauenheim}},\
  }\href@noop {} {\bibfield  {journal} {\bibinfo  {journal} {The Journal of
  Chemical Physics}\ }\textbf {\bibinfo {volume} {135}},\ \bibinfo {pages}
  {034701} (\bibinfo {year} {2011})}\BibitemShut {NoStop}%
\bibitem [{\citenamefont {Montero~de Hijes}\ \emph
  {et~al.}(2019{\natexlab{a}})\citenamefont {Montero~de Hijes}, \citenamefont
  {Espinosa}, \citenamefont {Vega},\ and\ \citenamefont
  {Sanz}}]{montero2019ice}%
  \BibitemOpen
  \bibfield  {author} {\bibinfo {author} {\bibfnamefont {P.}~\bibnamefont
  {Montero~de Hijes}}, \bibinfo {author} {\bibfnamefont {J.}~\bibnamefont
  {Espinosa}}, \bibinfo {author} {\bibfnamefont {C.}~\bibnamefont {Vega}}, \
  and\ \bibinfo {author} {\bibfnamefont {E.}~\bibnamefont {Sanz}},\ }\href@noop
  {} {\bibfield  {journal} {\bibinfo  {journal} {The Journal of Chemical
  Physics}\ }\textbf {\bibinfo {volume} {151}},\ \bibinfo {pages} {044509}
  (\bibinfo {year} {2019}{\natexlab{a}})}\BibitemShut {NoStop}%
\bibitem [{\citenamefont {Debenedetti}\ \emph {et~al.}(2020)\citenamefont
  {Debenedetti}, \citenamefont {Sciortino},\ and\ \citenamefont
  {Zerze}}]{debenedetti2020second}%
  \BibitemOpen
  \bibfield  {author} {\bibinfo {author} {\bibfnamefont {P.~G.}\ \bibnamefont
  {Debenedetti}}, \bibinfo {author} {\bibfnamefont {F.}~\bibnamefont
  {Sciortino}}, \ and\ \bibinfo {author} {\bibfnamefont {G.~H.}\ \bibnamefont
  {Zerze}},\ }\href@noop {} {\bibfield  {journal} {\bibinfo  {journal}
  {Science}\ }\textbf {\bibinfo {volume} {369}},\ \bibinfo {pages} {289}
  (\bibinfo {year} {2020})}\BibitemShut {NoStop}%
\bibitem [{\citenamefont {Lupi}\ \emph {et~al.}(2021)\citenamefont {Lupi},
  \citenamefont {V{\'a}zquez~Ram{\'\i}rez},\ and\ \citenamefont
  {Gallo}}]{lupi2021dynamical}%
  \BibitemOpen
  \bibfield  {author} {\bibinfo {author} {\bibfnamefont {L.}~\bibnamefont
  {Lupi}}, \bibinfo {author} {\bibfnamefont {B.}~\bibnamefont
  {V{\'a}zquez~Ram{\'\i}rez}}, \ and\ \bibinfo {author} {\bibfnamefont
  {P.}~\bibnamefont {Gallo}},\ }\href@noop {} {\bibfield  {journal} {\bibinfo
  {journal} {The Journal of Chemical Physics}\ }\textbf {\bibinfo {volume}
  {155}},\ \bibinfo {pages} {054502} (\bibinfo {year} {2021})}\BibitemShut
  {NoStop}%
\bibitem [{\citenamefont {Espinosa}\ \emph
  {et~al.}(2016{\natexlab{c}})\citenamefont {Espinosa}, \citenamefont {Vega},
  \citenamefont {Valeriani},\ and\ \citenamefont {Sanz}}]{espinosa2016seeding}%
  \BibitemOpen
  \bibfield  {author} {\bibinfo {author} {\bibfnamefont {J.~R.}\ \bibnamefont
  {Espinosa}}, \bibinfo {author} {\bibfnamefont {C.}~\bibnamefont {Vega}},
  \bibinfo {author} {\bibfnamefont {C.}~\bibnamefont {Valeriani}}, \ and\
  \bibinfo {author} {\bibfnamefont {E.}~\bibnamefont {Sanz}},\ }\href@noop {}
  {\bibfield  {journal} {\bibinfo  {journal} {The Journal of Chemical Physics}\
  }\textbf {\bibinfo {volume} {144}},\ \bibinfo {pages} {034501} (\bibinfo
  {year} {2016}{\natexlab{c}})}\BibitemShut {NoStop}%
\bibitem [{\citenamefont {Espinosa}\ \emph
  {et~al.}(2016{\natexlab{d}})\citenamefont {Espinosa}, \citenamefont {Vega},\
  and\ \citenamefont {Sanz}}]{espinosa2016ice}%
  \BibitemOpen
  \bibfield  {author} {\bibinfo {author} {\bibfnamefont {J.~R.}\ \bibnamefont
  {Espinosa}}, \bibinfo {author} {\bibfnamefont {C.}~\bibnamefont {Vega}}, \
  and\ \bibinfo {author} {\bibfnamefont {E.}~\bibnamefont {Sanz}},\ }\href@noop
  {} {\bibfield  {journal} {\bibinfo  {journal} {The Journal of Physical
  Chemistry C}\ }\textbf {\bibinfo {volume} {120}},\ \bibinfo {pages} {8068}
  (\bibinfo {year} {2016}{\natexlab{d}})}\BibitemShut {NoStop}%
\bibitem [{\citenamefont {Nos{\'e}}(1984)}]{nose1984unified}%
  \BibitemOpen
  \bibfield  {author} {\bibinfo {author} {\bibfnamefont {S.}~\bibnamefont
  {Nos{\'e}}},\ }\href@noop {} {\bibfield  {journal} {\bibinfo  {journal} {The
  Journal of Chemical Physics}\ }\textbf {\bibinfo {volume} {81}},\ \bibinfo
  {pages} {511} (\bibinfo {year} {1984})}\BibitemShut {NoStop}%
\bibitem [{\citenamefont {Hoover}(1985)}]{hoover1985canonical}%
  \BibitemOpen
  \bibfield  {author} {\bibinfo {author} {\bibfnamefont {W.~G.}\ \bibnamefont
  {Hoover}},\ }\href@noop {} {\bibfield  {journal} {\bibinfo  {journal}
  {Physical review A}\ }\textbf {\bibinfo {volume} {31}},\ \bibinfo {pages}
  {1695} (\bibinfo {year} {1985})}\BibitemShut {NoStop}%
\bibitem [{\citenamefont {Parrinello}\ and\ \citenamefont
  {Rahman}(1980)}]{parrinello1980crystal}%
  \BibitemOpen
  \bibfield  {author} {\bibinfo {author} {\bibfnamefont {M.}~\bibnamefont
  {Parrinello}}\ and\ \bibinfo {author} {\bibfnamefont {A.}~\bibnamefont
  {Rahman}},\ }\href@noop {} {\bibfield  {journal} {\bibinfo  {journal}
  {Physical Review Letters}\ }\textbf {\bibinfo {volume} {45}},\ \bibinfo
  {pages} {1196} (\bibinfo {year} {1980})}\BibitemShut {NoStop}%
\bibitem [{\citenamefont {Essmann}\ \emph {et~al.}(1995)\citenamefont
  {Essmann}, \citenamefont {Perera}, \citenamefont {Berkowitz}, \citenamefont
  {Darden}, \citenamefont {Lee},\ and\ \citenamefont
  {Pedersen}}]{essmann1995smooth}%
  \BibitemOpen
  \bibfield  {author} {\bibinfo {author} {\bibfnamefont {U.}~\bibnamefont
  {Essmann}}, \bibinfo {author} {\bibfnamefont {L.}~\bibnamefont {Perera}},
  \bibinfo {author} {\bibfnamefont {M.~L.}\ \bibnamefont {Berkowitz}}, \bibinfo
  {author} {\bibfnamefont {T.}~\bibnamefont {Darden}}, \bibinfo {author}
  {\bibfnamefont {H.}~\bibnamefont {Lee}}, \ and\ \bibinfo {author}
  {\bibfnamefont {L.~G.}\ \bibnamefont {Pedersen}},\ }\href@noop {} {\bibfield
  {journal} {\bibinfo  {journal} {The Journal of Chemical Physics}\ }\textbf
  {\bibinfo {volume} {103}},\ \bibinfo {pages} {8577} (\bibinfo {year}
  {1995})}\BibitemShut {NoStop}%
\bibitem [{\citenamefont {Bai}\ and\ \citenamefont {Li}(2005)}]{bai2005test}%
  \BibitemOpen
  \bibfield  {author} {\bibinfo {author} {\bibfnamefont {X.~M.}\ \bibnamefont
  {Bai}}\ and\ \bibinfo {author} {\bibfnamefont {M.}~\bibnamefont {Li}},\
  }\href@noop {} {\bibfield  {journal} {\bibinfo  {journal} {The Journal of
  Chemical Physics}\ }\textbf {\bibinfo {volume} {122}},\ \bibinfo {pages}
  {224510} (\bibinfo {year} {2005})}\BibitemShut {NoStop}%
\bibitem [{\citenamefont {Knott}\ \emph {et~al.}(2012)\citenamefont {Knott},
  \citenamefont {Molinero}, \citenamefont {Doherty},\ and\ \citenamefont
  {Peters}}]{knott2012homogeneous}%
  \BibitemOpen
  \bibfield  {author} {\bibinfo {author} {\bibfnamefont {B.~C.}\ \bibnamefont
  {Knott}}, \bibinfo {author} {\bibfnamefont {V.}~\bibnamefont {Molinero}},
  \bibinfo {author} {\bibfnamefont {M.~F.}\ \bibnamefont {Doherty}}, \ and\
  \bibinfo {author} {\bibfnamefont {B.}~\bibnamefont {Peters}},\ }\href@noop {}
  {\bibfield  {journal} {\bibinfo  {journal} {Journal of the American Chemical
  Society}\ }\textbf {\bibinfo {volume} {134}},\ \bibinfo {pages} {19544}
  (\bibinfo {year} {2012})}\BibitemShut {NoStop}%
\bibitem [{\citenamefont {Kelton}\ and\ \citenamefont
  {Greer}(2010)}]{kelton2010nucleation}%
  \BibitemOpen
  \bibfield  {author} {\bibinfo {author} {\bibfnamefont {K.~F.}\ \bibnamefont
  {Kelton}}\ and\ \bibinfo {author} {\bibfnamefont {A.~L.}\ \bibnamefont
  {Greer}},\ }\href@noop {} {\emph {\bibinfo {title} {Nucleation in condensed
  matter: applications in materials and biology}}}\ (\bibinfo  {publisher}
  {Elsevier},\ \bibinfo {year} {2010})\BibitemShut {NoStop}%
\bibitem [{\citenamefont {Kashchiev}(2000)}]{kashchiev2000nucleation}%
  \BibitemOpen
  \bibfield  {author} {\bibinfo {author} {\bibfnamefont {D.}~\bibnamefont
  {Kashchiev}},\ }\href@noop {} {\emph {\bibinfo {title} {Nucleation}}}\
  (\bibinfo  {publisher} {Elsevier},\ \bibinfo {year} {2000})\BibitemShut
  {NoStop}%
\bibitem [{\citenamefont {Kondo}(1956)}]{kondo1956thermodynamical}%
  \BibitemOpen
  \bibfield  {author} {\bibinfo {author} {\bibfnamefont {S.}~\bibnamefont
  {Kondo}},\ }\href@noop {} {\bibfield  {journal} {\bibinfo  {journal} {The
  Journal of Chemical Physics}\ }\textbf {\bibinfo {volume} {25}},\ \bibinfo
  {pages} {662} (\bibinfo {year} {1956})}\BibitemShut {NoStop}%
\bibitem [{\citenamefont {Rowlinson}\ and\ \citenamefont
  {Widom}(2013)}]{rowlinson2013molecular}%
  \BibitemOpen
  \bibfield  {author} {\bibinfo {author} {\bibfnamefont {J.~S.}\ \bibnamefont
  {Rowlinson}}\ and\ \bibinfo {author} {\bibfnamefont {B.}~\bibnamefont
  {Widom}},\ }\href@noop {} {\emph {\bibinfo {title} {Molecular theory of
  capillarity}}}\ (\bibinfo  {publisher} {Courier Corporation},\ \bibinfo
  {year} {2013})\BibitemShut {NoStop}%
\bibitem [{\citenamefont {Tr{\"o}ster}\ \emph {et~al.}(2012)\citenamefont
  {Tr{\"o}ster}, \citenamefont {Oettel}, \citenamefont {Block}, \citenamefont
  {Virnau},\ and\ \citenamefont {Binder}}]{troster2012numerical}%
  \BibitemOpen
  \bibfield  {author} {\bibinfo {author} {\bibfnamefont {A.}~\bibnamefont
  {Tr{\"o}ster}}, \bibinfo {author} {\bibfnamefont {M.}~\bibnamefont {Oettel}},
  \bibinfo {author} {\bibfnamefont {B.}~\bibnamefont {Block}}, \bibinfo
  {author} {\bibfnamefont {P.}~\bibnamefont {Virnau}}, \ and\ \bibinfo {author}
  {\bibfnamefont {K.}~\bibnamefont {Binder}},\ }\href@noop {} {\bibfield
  {journal} {\bibinfo  {journal} {The Journal of Chemical Physics}\ }\textbf
  {\bibinfo {volume} {136}},\ \bibinfo {pages} {064709} (\bibinfo {year}
  {2012})}\BibitemShut {NoStop}%
\bibitem [{\citenamefont {Montero~de Hijes}\ \emph {et~al.}(2020)\citenamefont
  {Montero~de Hijes}, \citenamefont {Shi}, \citenamefont {Noya}, \citenamefont
  {Santiso}, \citenamefont {Gubbins}, \citenamefont {Sanz},\ and\ \citenamefont
  {Vega}}]{montero2020young}%
  \BibitemOpen
  \bibfield  {author} {\bibinfo {author} {\bibfnamefont {P.}~\bibnamefont
  {Montero~de Hijes}}, \bibinfo {author} {\bibfnamefont {K.}~\bibnamefont
  {Shi}}, \bibinfo {author} {\bibfnamefont {E.~G.}\ \bibnamefont {Noya}},
  \bibinfo {author} {\bibfnamefont {E.}~\bibnamefont {Santiso}}, \bibinfo
  {author} {\bibfnamefont {K.}~\bibnamefont {Gubbins}}, \bibinfo {author}
  {\bibfnamefont {E.}~\bibnamefont {Sanz}}, \ and\ \bibinfo {author}
  {\bibfnamefont {C.}~\bibnamefont {Vega}},\ }\href@noop {} {\bibfield
  {journal} {\bibinfo  {journal} {The Journal of Chemical Physics}\ }\textbf
  {\bibinfo {volume} {153}},\ \bibinfo {pages} {191102} (\bibinfo {year}
  {2020})}\BibitemShut {NoStop}%
\bibitem [{\citenamefont {Kashchiev}(2020)}]{kashchiev2020nucleation}%
  \BibitemOpen
  \bibfield  {author} {\bibinfo {author} {\bibfnamefont {D.}~\bibnamefont
  {Kashchiev}},\ }\href@noop {} {\bibfield  {journal} {\bibinfo  {journal} {The
  Journal of Chemical Physics}\ }\textbf {\bibinfo {volume} {153}},\ \bibinfo
  {pages} {124509} (\bibinfo {year} {2020})}\BibitemShut {NoStop}%
\bibitem [{\citenamefont {Montero~de Hijes}\ and\ \citenamefont
  {Vega}(2022)}]{montero2022thermodynamics}%
  \BibitemOpen
  \bibfield  {author} {\bibinfo {author} {\bibfnamefont {P.}~\bibnamefont
  {Montero~de Hijes}}\ and\ \bibinfo {author} {\bibfnamefont {C.}~\bibnamefont
  {Vega}},\ }\href@noop {} {\bibfield  {journal} {\bibinfo  {journal} {The
  Journal of Chemical Physics}\ }\textbf {\bibinfo {volume} {156}},\ \bibinfo
  {pages} {014505} (\bibinfo {year} {2022})}\BibitemShut {NoStop}%
\bibitem [{\citenamefont {Montero~de Hijes}\ \emph
  {et~al.}(2019{\natexlab{b}})\citenamefont {Montero~de Hijes}, \citenamefont
  {Espinosa}, \citenamefont {Sanz},\ and\ \citenamefont
  {Vega}}]{montero2019interfacial}%
  \BibitemOpen
  \bibfield  {author} {\bibinfo {author} {\bibfnamefont {P.}~\bibnamefont
  {Montero~de Hijes}}, \bibinfo {author} {\bibfnamefont {J.~R.}\ \bibnamefont
  {Espinosa}}, \bibinfo {author} {\bibfnamefont {E.}~\bibnamefont {Sanz}}, \
  and\ \bibinfo {author} {\bibfnamefont {C.}~\bibnamefont {Vega}},\ }\href@noop
  {} {\bibfield  {journal} {\bibinfo  {journal} {The Journal of Chemical
  Physics}\ }\textbf {\bibinfo {volume} {151}},\ \bibinfo {pages} {144501}
  (\bibinfo {year} {2019}{\natexlab{b}})}\BibitemShut {NoStop}%
\bibitem [{\citenamefont {Espinosa}\ \emph {et~al.}(2017)\citenamefont
  {Espinosa}, \citenamefont {Soria}, \citenamefont {Ramirez}, \citenamefont
  {Valeriani}, \citenamefont {Vega},\ and\ \citenamefont
  {Sanz}}]{espinosa2017role}%
  \BibitemOpen
  \bibfield  {author} {\bibinfo {author} {\bibfnamefont {J.~R.}\ \bibnamefont
  {Espinosa}}, \bibinfo {author} {\bibfnamefont {G.~D.}\ \bibnamefont {Soria}},
  \bibinfo {author} {\bibfnamefont {J.}~\bibnamefont {Ramirez}}, \bibinfo
  {author} {\bibfnamefont {C.}~\bibnamefont {Valeriani}}, \bibinfo {author}
  {\bibfnamefont {C.}~\bibnamefont {Vega}}, \ and\ \bibinfo {author}
  {\bibfnamefont {E.}~\bibnamefont {Sanz}},\ }\href@noop {} {\bibfield
  {journal} {\bibinfo  {journal} {The Journal of Physical Chemistry Letters}\
  }\textbf {\bibinfo {volume} {8}},\ \bibinfo {pages} {4486} (\bibinfo {year}
  {2017})}\BibitemShut {NoStop}%
\bibitem [{\citenamefont {Lechner}\ and\ \citenamefont
  {Dellago}(2008)}]{lechner2008accurate}%
  \BibitemOpen
  \bibfield  {author} {\bibinfo {author} {\bibfnamefont {W.}~\bibnamefont
  {Lechner}}\ and\ \bibinfo {author} {\bibfnamefont {C.}~\bibnamefont
  {Dellago}},\ }\href@noop {} {\bibfield  {journal} {\bibinfo  {journal} {The
  Journal of Chemical Physics}\ }\textbf {\bibinfo {volume} {129}},\ \bibinfo
  {pages} {114707} (\bibinfo {year} {2008})}\BibitemShut {NoStop}%
\bibitem [{\citenamefont {Vega}\ \emph {et~al.}(2008)\citenamefont {Vega},
  \citenamefont {Sanz}, \citenamefont {Abascal},\ and\ \citenamefont
  {Noya}}]{vega2008determination}%
  \BibitemOpen
  \bibfield  {author} {\bibinfo {author} {\bibfnamefont {C.}~\bibnamefont
  {Vega}}, \bibinfo {author} {\bibfnamefont {E.}~\bibnamefont {Sanz}}, \bibinfo
  {author} {\bibfnamefont {J.}~\bibnamefont {Abascal}}, \ and\ \bibinfo
  {author} {\bibfnamefont {E.}~\bibnamefont {Noya}},\ }\href@noop {} {\bibfield
   {journal} {\bibinfo  {journal} {Journal of Physics: Condensed Matter}\
  }\textbf {\bibinfo {volume} {20}},\ \bibinfo {pages} {153101} (\bibinfo
  {year} {2008})}\BibitemShut {NoStop}%
\bibitem [{\citenamefont {Schmelzer}\ \emph {et~al.}(2019)\citenamefont
  {Schmelzer}, \citenamefont {Abyzov},\ and\ \citenamefont
  {Baidakov}}]{schmelzer2019entropy}%
  \BibitemOpen
  \bibfield  {author} {\bibinfo {author} {\bibfnamefont {J.~W.}\ \bibnamefont
  {Schmelzer}}, \bibinfo {author} {\bibfnamefont {A.~S.}\ \bibnamefont
  {Abyzov}}, \ and\ \bibinfo {author} {\bibfnamefont {V.~G.}\ \bibnamefont
  {Baidakov}},\ }\href@noop {} {\bibfield  {journal} {\bibinfo  {journal}
  {Entropy}\ }\textbf {\bibinfo {volume} {21}},\ \bibinfo {pages} {670}
  (\bibinfo {year} {2019})}\BibitemShut {NoStop}%
\bibitem [{\citenamefont {Baidakov}\ and\ \citenamefont
  {Protsenko}(2019)}]{baidakov2019spontaneous}%
  \BibitemOpen
  \bibfield  {author} {\bibinfo {author} {\bibfnamefont {V.~G.}\ \bibnamefont
  {Baidakov}}\ and\ \bibinfo {author} {\bibfnamefont {K.~R.}\ \bibnamefont
  {Protsenko}},\ }\href@noop {} {\bibfield  {journal} {\bibinfo  {journal} {The
  Journal of Physical Chemistry B}\ }\textbf {\bibinfo {volume} {123}},\
  \bibinfo {pages} {8103} (\bibinfo {year} {2019})}\BibitemShut {NoStop}%
\bibitem [{\citenamefont {Molinero}\ and\ \citenamefont
  {Moore}(2009)}]{molinero2009water}%
  \BibitemOpen
  \bibfield  {author} {\bibinfo {author} {\bibfnamefont {V.}~\bibnamefont
  {Molinero}}\ and\ \bibinfo {author} {\bibfnamefont {E.~B.}\ \bibnamefont
  {Moore}},\ }\href@noop {} {\bibfield  {journal} {\bibinfo  {journal} {The
  Journal of Physical Chemistry B}\ }\textbf {\bibinfo {volume} {113}},\
  \bibinfo {pages} {4008} (\bibinfo {year} {2009})}\BibitemShut {NoStop}%
\bibitem [{\citenamefont {Handel}\ \emph {et~al.}(2008)\citenamefont {Handel},
  \citenamefont {Davidchack}, \citenamefont {Anwar},\ and\ \citenamefont
  {Brukhno}}]{handel2008direct}%
  \BibitemOpen
  \bibfield  {author} {\bibinfo {author} {\bibfnamefont {R.}~\bibnamefont
  {Handel}}, \bibinfo {author} {\bibfnamefont {R.~L.}\ \bibnamefont
  {Davidchack}}, \bibinfo {author} {\bibfnamefont {J.}~\bibnamefont {Anwar}}, \
  and\ \bibinfo {author} {\bibfnamefont {A.}~\bibnamefont {Brukhno}},\
  }\href@noop {} {\bibfield  {journal} {\bibinfo  {journal} {Physical Review
  Letters}\ }\textbf {\bibinfo {volume} {100}},\ \bibinfo {pages} {036104}
  (\bibinfo {year} {2008})}\BibitemShut {NoStop}%
\bibitem [{\citenamefont {Davidchack}\ \emph {et~al.}(2012)\citenamefont
  {Davidchack}, \citenamefont {Handel}, \citenamefont {Anwar},\ and\
  \citenamefont {Brukhno}}]{davidchack2012ice}%
  \BibitemOpen
  \bibfield  {author} {\bibinfo {author} {\bibfnamefont {R.~L.}\ \bibnamefont
  {Davidchack}}, \bibinfo {author} {\bibfnamefont {R.}~\bibnamefont {Handel}},
  \bibinfo {author} {\bibfnamefont {J.}~\bibnamefont {Anwar}}, \ and\ \bibinfo
  {author} {\bibfnamefont {A.~V.}\ \bibnamefont {Brukhno}},\ }\href@noop {}
  {\bibfield  {journal} {\bibinfo  {journal} {Journal of Chemical Theory and
  Computation}\ }\textbf {\bibinfo {volume} {8}},\ \bibinfo {pages} {2383}
  (\bibinfo {year} {2012})}\BibitemShut {NoStop}%
\bibitem [{\citenamefont {Rozmanov}\ and\ \citenamefont
  {Kusalik}(2012)}]{rozmanov2012anisotropy}%
  \BibitemOpen
  \bibfield  {author} {\bibinfo {author} {\bibfnamefont {D.}~\bibnamefont
  {Rozmanov}}\ and\ \bibinfo {author} {\bibfnamefont {P.~G.}\ \bibnamefont
  {Kusalik}},\ }\href@noop {} {\bibfield  {journal} {\bibinfo  {journal} {The
  Journal of chemical physics}\ }\textbf {\bibinfo {volume} {137}},\ \bibinfo
  {pages} {094702} (\bibinfo {year} {2012})}\BibitemShut {NoStop}%
\bibitem [{\citenamefont {Gibbs}(1928)}]{gibbs1928collected}%
  \BibitemOpen
  \bibfield  {author} {\bibinfo {author} {\bibfnamefont {J.~W.}\ \bibnamefont
  {Gibbs}},\ }\href@noop {} {\emph {\bibinfo {title} {The collected works of J.
  Willard Gibbs, volume I: thermodynamics}}}\ (\bibinfo  {publisher} {Yale
  University Press},\ \bibinfo {year} {1928})\BibitemShut {NoStop}%
\bibitem [{\citenamefont {Qiu}\ \emph {et~al.}(2018)\citenamefont {Qiu},
  \citenamefont {Lupi},\ and\ \citenamefont {Molinero}}]{qiu2018water}%
  \BibitemOpen
  \bibfield  {author} {\bibinfo {author} {\bibfnamefont {Y.}~\bibnamefont
  {Qiu}}, \bibinfo {author} {\bibfnamefont {L.}~\bibnamefont {Lupi}}, \ and\
  \bibinfo {author} {\bibfnamefont {V.}~\bibnamefont {Molinero}},\ }\href@noop
  {} {\bibfield  {journal} {\bibinfo  {journal} {The Journal of Physical
  Chemistry B}\ }\textbf {\bibinfo {volume} {122}},\ \bibinfo {pages} {3626}
  (\bibinfo {year} {2018})}\BibitemShut {NoStop}%
\bibitem [{\citenamefont {Eisenberg}\ and\ \citenamefont
  {Kauzmann}(2005)}]{eisenberg2005structure}%
  \BibitemOpen
  \bibfield  {author} {\bibinfo {author} {\bibfnamefont {D.}~\bibnamefont
  {Eisenberg}}\ and\ \bibinfo {author} {\bibfnamefont {W.}~\bibnamefont
  {Kauzmann}},\ }\href@noop {} {\emph {\bibinfo {title} {The structure and
  properties of water}}}\ (\bibinfo  {publisher} {OUP Oxford},\ \bibinfo {year}
  {2005})\BibitemShut {NoStop}%
\bibitem [{\citenamefont {Abascal}\ \emph {et~al.}(2009)\citenamefont
  {Abascal}, \citenamefont {Sanz},\ and\ \citenamefont
  {Vega}}]{abascal2009triple}%
  \BibitemOpen
  \bibfield  {author} {\bibinfo {author} {\bibfnamefont {J.~L.}\ \bibnamefont
  {Abascal}}, \bibinfo {author} {\bibfnamefont {E.}~\bibnamefont {Sanz}}, \
  and\ \bibinfo {author} {\bibfnamefont {C.}~\bibnamefont {Vega}},\ }\href@noop
  {} {\bibfield  {journal} {\bibinfo  {journal} {Physical Chemistry Chemical
  Physics}\ }\textbf {\bibinfo {volume} {11}},\ \bibinfo {pages} {556}
  (\bibinfo {year} {2009})}\BibitemShut {NoStop}%
\end{thebibliography}%

\end{document}